\documentclass[sigplan,10pt]{acmart}

\AtBeginDocument{%
  }

\usepackage{tikz}
\usepackage{amsmath}
\usepackage{xspace}
\usepackage{enumitem}
\usepackage{wrapfig}
\usepackage{float}
\usepackage[font=small]{caption}
\usepackage{subcaption}
\usepackage{booktabs}
\usepackage{threeparttable}

\usepackage[ruled, linesnumbered, vlined]{algorithm2e}
\usepackage{amsfonts}
\usetikzlibrary{calc}
\usepackage{multirow}
\usepackage{libertine}

\newcommand{\EdgeFlow}{\textsf{\textsc{EdgeFlow}}\xspace}

\newcommand{\ie}{{\em i.e.},\xspace}
\newcommand{\eg}{{\em e.g.},\xspace}

\usepackage{url}
\usepackage{xurl}

\usepackage{listings}
\usepackage{color}
\definecolor{darkgreen}{RGB}{0, 102, 5}

\renewcommand\footnotetextcopyrightpermission[1]{}
\acmSubmissionID{884}

\begin{document}

\title{\EdgeFlow: Fast Cold Starts for LLMs on Mobile Devices}

\author{Yongsheng Yan}
\affiliation{%
  \institution{College of Computer Science and Artificial Intelligence, Fudan University}
  \country{China}
}
\email{ysyan24@m.fudan.edu.cn}
\author{Jiacheng Shen}
\affiliation{%
  \institution{Duke Kunshan University}
  \country{China}
}
\email{jc.shen@dukekunshan.edu.cn}
\author{Xuchuan Luo}
\affiliation{%
  \institution{College of Computer Science and Artificial Intelligence, Fudan University}
  \country{China}
}
\email{xcluo23@m.fudan.edu.cn}
\author{Yangfan Zhou}
\affiliation{%
  \institution{College of Computer Science and Artificial Intelligence, Fudan University}
  \country{China}
}
\email{zyf@fudan.edu.cn}

\begin{abstract}
Deploying large language models (LLMs) on mobile devices is an emerging trend to enable data privacy and offline accessibility of LLM applications.
Modern mobile neural processing units (NPUs) make such deployment increasingly feasible.
However, existing mobile LLM inference frameworks suffer from high start-up latency due to their inevitable cold starts, \ie launching LLM inferences when the model is not hosted in device memory.
In this paper, we identify the key bottleneck of mobile LLM cold starts as the waste of flash bandwidth on unimportant model parameters.
We design \textbf{\EdgeFlow}, a mobile LLM inference framework that mitigates the cold start issue by adaptively adjusting the precisions of LLM parameters.
Specifically, \EdgeFlow leverages 
1) an \textit{NPU-aware adaptive quantization} algorithm that assigns different precisions to weights in a finer granularity according to their importance and NPU constraints,
2) an \textit{SIMD-friendly packing format} that accelerates the transformation of various-precision weights into fixed-sized NPU-native data types, and
3) a \textit{synergistic granular pipeline} that coordinates CPU and NPU computation in a fine-grained and dynamic manner.
Experimental results show that \EdgeFlow reduces cold-start latency by up to $4.07\times$ compared with three state-of-the-art mobile LLM inference frameworks, \ie llama.cpp, MNN, and llm.npu, under comparable model accuracy.
\end{abstract}
\maketitle
\makeatletter
\def\@oddhead{}
\def\@evenhead{}
\makeatother
\section{Introduction}
\label{sec:introduction}
\noindent
With the increasing adoption of large language models (LLMs) in everyday applications, \eg personal assistants~\cite{arxiv24mobile-agent,nips24mobile-agent2,arxiv25mobi-agent}, smart home control~\cite{mobubi24sasha,acl2025home-assistant}, and healthcare~\cite{arxiv24phllm,bhi24phisiollm}, there is a growing trend towards deploying LLM inference directly on mobile devices to ensure data privacy and 
offline accessibility~\cite{llama-cpp,arxiv24transformer-lite,cvpr25bluelm}.
Many LLM inference frameworks are proposed to improve the execution efficiency of LLMs on resource-constrained mobile devices~\cite{asplos25mllm,sosp25heteroinfer,mmasia24mnn}. 

However, the \textit{\textbf{cold start}} issue is largely overlooked in existing mobile LLM inference frameworks.
Cold start is a critical performance bottleneck for mobile applications~\cite{fast11fast,atc21asap,fast23Paralfetch,fast25archer,atc25prm,hpca25ariadne,atc20acclaim}.
It happens when an application is invoked, but the required data are not loaded into device memory.
This issue is exacerbated in mobile LLM applications, \eg conversational agents and UI automation tasks~\cite{nips23chatbot-arena,mobicom24droidtask}, where a huge amount of model weights have to be loaded into memory.
Our preliminary experiments show that llm.npu~\cite{asplos25mllm}, the state-of-the-art mobile LLM inference framework, suffers from a 9.1-second time to first token (TTFT) during a cold start of executing the Llama3 8B~\cite{Llama3-8B} model under 128 input tokens.
Such latency is an order of magnitude greater than the cold start of conventional mobile applications~\cite{mobicom26appflow,mobicom23swam}, which significantly degrade user experiences~\cite{nielsen1993response,chiir21response}.
Even worse, cold starts of LLM frameworks frequently occur since mobile operating systems can not always keep the massive model parameters in limited device memory~\cite{asplos24moreapps,android_lmkd,apple_memory_pressure}.

In this paper, we identify the primary cause of long cold-start latency as suboptimal utilization of flash bandwidth when loading LLMs.
Specifically, existing LLM inference frameworks typically quantize model weights into lower precisions to reduce model sizes~\cite{iclr23gptq,icml23smoothquant}.
However, conventional quantization techniques uniformly quantize all weights into the same precision, \eg INT8, to align with hardware-supported data types.
Such an approach overlooks that different weights have different contributions to model accuracy~\cite{mlsys24awq}.
Flash bandwidths are hence wasted on transferring unimportant weights that are under-quantized.

To address this problem, we propose to employ adaptive quantization to achieve a better trade-off between model accuracy and flash bandwidth utilization.
The key idea is to determine and assign the required precision of each weight based on its importance to model accuracy.
Nevertheless, this approach presents several challenges that call for a co-design of inference systems and quantization algorithms.

\textbf{\textit{(1) Precision assignments under NPU constraints.}}
Existing mobile LLM inference frameworks leverage NPUs to accelerate inference efficiency~\cite{asplos25mllm,arxiv24powerinfer2,sosp25heteroinfer,eurosys26llamanpu}.
Production mobile NPUs, \eg Qualcomm Hexagon NPU~\cite{hexagonnpu}, only accelerate matrix multiplications (matmuls) for tensors under uniform and coarse-grained quantization.
Such constraints impede the adaptive assignment of precisions to individual weights and hinder the execution of existing importance-aware quantization methods that adopt non-uniform, fine-grained quantization on CPUs and GPUs~\cite{mlsys24awq,arxiv25cmpq}.

\textbf{\textit{(2) Costly unpacking for variable-precision weights.}}
Low-bit quantized model weights have to be unpacked into NPU-native data types, \eg INT8 and FP16, by the CPU before computation.
Existing approaches store quantized models in an unpack-friendly format to reduce the computational overhead during unpacking~\cite{llama-cpp,mlsys24awq}. 
However, these formats are designed for weights quantized with the same precision.
Naively storing variable-precision weights in these formats incurs significant overhead due to the excessive memory accesses during unpacking computation.

\textbf{\textit{(3) Imbalanced workloads between CPU and NPU.}}
Unpacking weights and conducting prefill computation require the collaboration between the CPU and NPU.
Existing approaches schedule computational tasks to the two processors in a coarse-grained and static manner~\cite{asplos25mllm,arxiv24powerinfer2}.
Unfortunately, such scheduling results in significant bubbles in the computation pipeline.
This would make the long computation time another bottleneck for cold-starts.

We design \EdgeFlow, a system that co-designs a quantization algorithm with system-level optimizations to reduce cold-start latencies for mobile LLM inferences.
First, we design an \textit{NPU-aware adaptive quantization} algorithm to differentiate the importance of weights and assign fine-grained precisions under NPU constraints, enabling fast model loading with comparable accuracy.
Second, we develop an \textit{SIMD-friendly packing format} with an SIMD-optimized unpacking algorithm to facilitate rapid transformation of adaptively quantized weights into NPU-native data types.
Finally, we propose a \textit{synergistic granular pipeline} that balances CPU and NPU computation in a dynamic, fine-grained manner to maximize compute resource utilization during cold starts.

We implement \EdgeFlow from scratch and evaluate it with various LLMs~\cite{Llama3-8B, Mistral-7B, Phi3-3.8B, Qwen1.5-1.8B} and datasets~\cite{acl16lambada,aaai20windogrande,acl18obqa,iclr21mmlu,acl19hellaswag,iclr17wikitext2}.
We compare \EdgeFlow with two open-source approaches, \ie llama.cpp~\cite{llama-cpp} and MNN~\cite{mmasia24mnn}, and a state-of-the-art NPU-accelerated mobile LLM inference framework, \ie llm.npu~\cite{asplos25mllm}.
Our evaluation results show that with the same model accuracy, \EdgeFlow reduces cold-start latency, \ie TTFT, by $4.07\times$ and $2.40\times$ compared with llama.cpp and MNN.
Compared with llm.npu, \EdgeFlow achieves $1.55\times$ speedup in TTFT and up to $2.52\%$ improvements in accuracy.
We plan to open source \EdgeFlow in the near future.

Our contributions are summarized as follows:
\begin{itemize}[noitemsep, topsep=0pt, parsep=0pt, partopsep=0pt]
    \item We break down the cold-start issue of mobile LLM deployments, and show that the inefficient flash bandwidth utilization significantly deteriorates TTFT.
    \item We propose \EdgeFlow that co-designs the NPU-aware adaptive quantization algorithm with the SIMD-friendly packing format and the synergistic granular pipeline to collaboratively reduce cold-start latencies.
    \item We implement \EdgeFlow and show that it significantly reduces cold-start latency while preserving inference accuracy under multiple models and workloads.
\end{itemize}
\section{Background}\label{sec:background}
\subsection{Mobile NPUs for LLM Inference}
\noindent 
NPUs are widely adopted in smartphones due to their energy-efficient computing capabilities under AI workloads, \eg Qualcomm Hexagon NPU~\cite{hexagonnpu} and Apple Neural Engine~\cite{ane}.
In this paper, we focus on system design with Qualcomm Hexagon NPU, which is one of the most widely adopted NPUs on Android phones.
The proposed techniques can also be adapted to other platforms with engineering efforts.

Hexagon NPU provides developers with QNN SDK~\cite{qnn} to leverage various hardware accelerators inside NPUs.
QNN offers a toolchain to quantize and compile models into NPU executables, and a runtime to efficiently execute models over NPUs.
Specifically, to deploy a model on NPU, users first need to assemble a computation graph with high-level operators provided by QNN.
The computation graph is then compiled into an execution graph, \ie NPU executables, before being loaded into NPU with QNN runtime APIs.
During the whole process, the computation graph compilation is the most time-consuming step since QNN needs to conduct a series of NPU-specific optimizations, \eg fusing operators, scheduling tiles and loops, and reordering weights according to the memory layout of NPUs~\cite{asplos25mllm,eurosys26llamanpu}.
 
\begin{table}[t!]
    \centering
    \small
    \caption{Existing post-training quantization approaches.}
    \vspace{-6pt}
    \begin{tabular}{l|l}
    \toprule
    \textbf{Dimensions}   & \textbf{Categories} \\
    \midrule
    \textbf{Target}       & {Weight Tensors} / {Activation Tensors} \\
    \textbf{Timing}       & {Static} / {Dynamic} \\
    \textbf{Mapping Type} & {Uniform} / {Non-uniform} \\
    \textbf{Granularity}  & {Per-tensor} / {Per-channel} / {Per-block} \\
    \bottomrule
    \end{tabular}
    \label{tab:quantization-overview}
\end{table}
\subsection{Post-Training Quantization}
\noindent
Post-training quantization is widely adopted in mobile LLM deployments to reduce model sizes and runtime memory footprint.
The key idea is to map tensors into low-precision data types with some metadata and dequantize, \ie map back to high-precision tensors, during or after computation to maintain model accuracy.
As shown in Table~\ref{tab:quantization-overview}, existing quantization techniques can be classified in terms of quantization target, timing, mapping types, and granularity.
In terms of \textit{target}, there are two types of tensors during LLM inference, \ie weight tensors storing static model parameters and activation tensors storing intermediate results, \eg KV caches.
Both types of tensors are considered as targets for quantization.
In terms of \textit{timing}, static approaches quantize tensors offline, while dynamic approaches quantize them during inference.
In terms of \textit{mapping types}, uniform approaches quantize tensors with linear transformations, \ie one high-precision value in a tensor can be mapped to a low-precision one with a scale and a bias.
Uniform quantization can be further classified into symmetric and asymmetric methods based on whether the bias is $0$ or not.
Non-uniform approaches quantize tensors with non-linear transformations, \eg codebooks~\cite{emnlp24vptq}.
Finally, the \textit{granularity} refers to how many weights share the same mapping metadata.
Per-tensor approaches share metadata for an entire tensor, per-channel ones share metadata for a row or a column, and per-block ones share metadata among fixed-size groups.

Existing quantization schemes quantize tensors into a single precision~\cite{iclr23gptq,mlsys24awq,nips24quarot} or a limited group of precisions, \eg INT4 and INT8~\cite{arxiv25cmpq,nips23llm-mq}.
This results in inefficient bandwidth utilization during cold starts.
The essential reason is that mobile accelerators only contain arithmetic units to compute tensors with a limited set of data types, \eg INT8, INT16, and INT32. 
Mixed-precision tensors have to be transformed into tensors with native data types before computation, which incurs additional transformation overhead.

\section{Motivation and Challenges}
\label{sec:analysis}
\begin{figure}[t]
    \centering
    \begin{tikzpicture}
        \node[anchor=south west, inner sep=0] (img)
            {\includegraphics[width=1\linewidth]{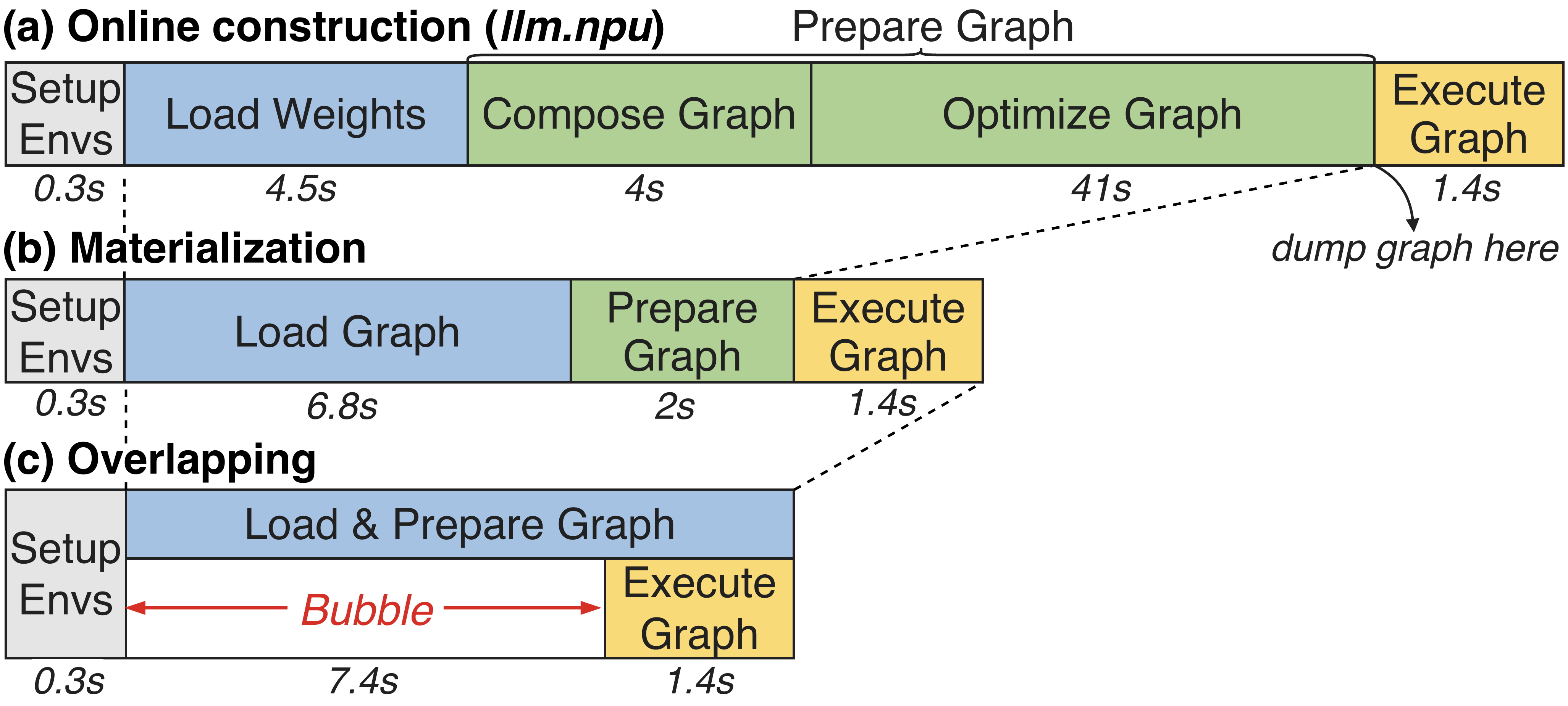}};
        \phantomsubcaption
        \label{fig:cold_start_process:a}
        \phantomsubcaption
        \label{fig:cold_start_process:b}
        \phantomsubcaption
        \label{fig:cold_start_process:c}
    \end{tikzpicture}
    \caption{Breakdown of the cold-start latencies of llm.npu and two straightforward optimizations, \ie materialization and overlapping.}
    \label{fig:cold_start_process}
    \vspace{-0.1in}
\end{figure}

\begin{figure}[t]
    \centering
    \includegraphics[width=1\linewidth]{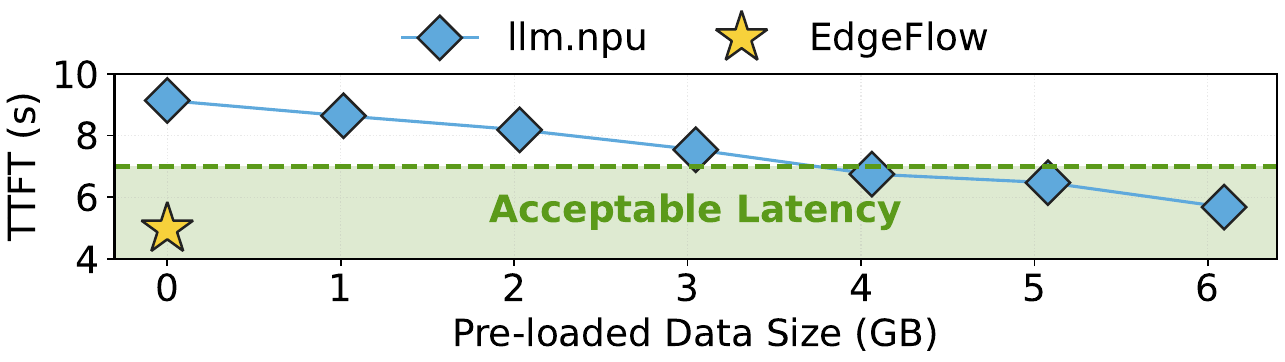}
    \vspace{-6mm}
    \caption{TTFT v.s. pre-loaded data size for llm.npu, including user-acceptable latency region and EdgeFlow measurements.}
    \label{fig:motivation_preload}
    \vspace{-0.1in}
\end{figure}

\noindent 
This section first breaks down the cold-start process with both theoretical and experimental analyses, motivating the need to adopt adaptive quantization (\S~\ref{sec:motivation}).
We then discuss the challenges of integrating adaptive quantization to mobile LLM inference systems (\S~\ref{sec:challenge}).
All experiments are conducted on a Xiaomi 15 Pro with Hexagon NPU, 16 GB of RAM, and 512 GB of flash storage.

\subsection{Analysis of the Cold-Start Issue}\label{sec:motivation}

\noindent
We use llm.npu~\cite{asplos25mllm}, the state-of-the-art NPU-based mobile LLM inference framework, to showcase the cold start process of NPU-based mobile LLM inference.
As shown in Figure~\ref{fig:cold_start_process:a}, the cold start process of llm.npu can be summarized into $4$ steps, \ie setup environments, load weights, prepare execution graphs, and execute.
According to our experiments on Llama3 8B with a 128-token prompt, llm.npu exhibits a cold-start latency of up to 51.2 seconds.
Its cold-start latency is mainly dominated by the overhead of on-the-fly graph preparation and weight loading.

Two methods can be adopted to reduce the pure cold-start latency, \ie materialization and overlap.
First, materialization stores the optimized computation graph offline and directly loads the graph during cold starts.
As shown in Figure~\ref{fig:cold_start_process:b}, such an approach eliminates the graph optimization overhead from the critical path and reduces cold-start latency from 51.2s to 10.5s.
Second, we can overlap graph loading with prefill computation between consecutive layers by decomposing the graph into multiple subgraphs, one for each layer.
Figure~\ref{fig:cold_start_process:c} illustrates that the overlapping method further reduces the cold-start latency to 9.1s.
Unfortunately, even with these optimizations, the latency remains unacceptable for interactive mobile applications, as users typically lose patience if the wait time exceeds 7s~\cite{chiir21response}.

The key bottleneck of existing approaches lies in the weight loading overhead, as indicated by the 7.4s pipeline bubble.
Many mobile systems prefetch application data from flash to reduce data loading latency during cold starts~\cite{atc21asap,fast23Paralfetch,mobicom26appflow,mobisys12falcon}.
However, as shown in Figure~\ref{fig:motivation_preload}, 4 GB of data must be prefetched and reserved in memory to achieve a satisfactory cold-start latency.
The consumed memory accounts for 25\% of total memory in our high-end testbed, which is a significant overhead for resource-constrained mobile devices.

In contrast, \EdgeFlow focuses on reducing the pure cold start latency, which can meet the latency requirement without reserving memory in advance.
To achieve this, we propose an adaptive quantization strategy to relieve the flash bandwidth bottleneck on the critical path of cold starts.
Our opportunity is that not all weights are equally important to the model accuracy~\cite{mlsys24awq,arxiv25cmpq,nips23llm-mq}.
Less precision, \ie bits, can be assigned to unimportant weights to optimize flash bandwidth utilization.

\subsection{Challenges of Adaptive Quantization}\label{sec:challenge}

\noindent 
In this section, we introduce three challenges when applying adaptive quantization in a mobile LLM inference system.

\textbf{Challenge 1: \textit{Precision assignments under NPU constraints.}}
NPUs only accelerate a limited set of quantized matmuls, which imposes two constraints on quantization algorithms.
First, all tensors, \ie weight and activation tensors, have to be quantized in a \textit{static, uniform, and symmetric} manner.
Second, activations can only be quantized in \textit{per-tensor} granularity, while weights can be quantized in \textit{per-channel} granularity only on output channels, \ie columns.
Such constraints result in poor matmul efficiency for tensors quantized by existing importance-aware quantization approaches since they often require more fine-grained quantization.

\begin{table}[t!]
    \centering
    \small
    \caption{Comparison between NPU constraints and existing importance-aware quantization approaches.}
    \begin{tabular}{
        p{0.180\linewidth}|
        p{0.132\linewidth}
        p{0.250\linewidth}
        p{0.258\linewidth}
    }
    \toprule
    \textbf{Method} 
        & \textbf{Target} 
        & \textbf{Mapping Type} 
        & \textbf{Granularity} \\
    \midrule
    \textbf{AWQ}~\cite{mlsys24awq} 
        & Weights 
        & Uniform (Asym.)
        & Per-block \\
    \textbf{CMPQ}~\cite{arxiv25cmpq} 
        & Weights 
        & Non-uniform 
        & Per-channel (in) \\
    \midrule
        \multirow{2}{*}{
            \hspace{-1.5mm}
            \begin{minipage}{0.2\linewidth}
            \raggedright \textbf{NPU \\ Constraints}
            \end{minipage}
        }
        & Weights 
        & \multirow{2}{*}{Uniform (Sym.)}
        & Per-channel (out) \\
        & Activations
        & 
        & Per-tensor \\
    \bottomrule
    \end{tabular}
    \label{tab:challenge1-constraint}
\end{table}

\begin{figure}[t]
    \centering
    \includegraphics[width=1\columnwidth]{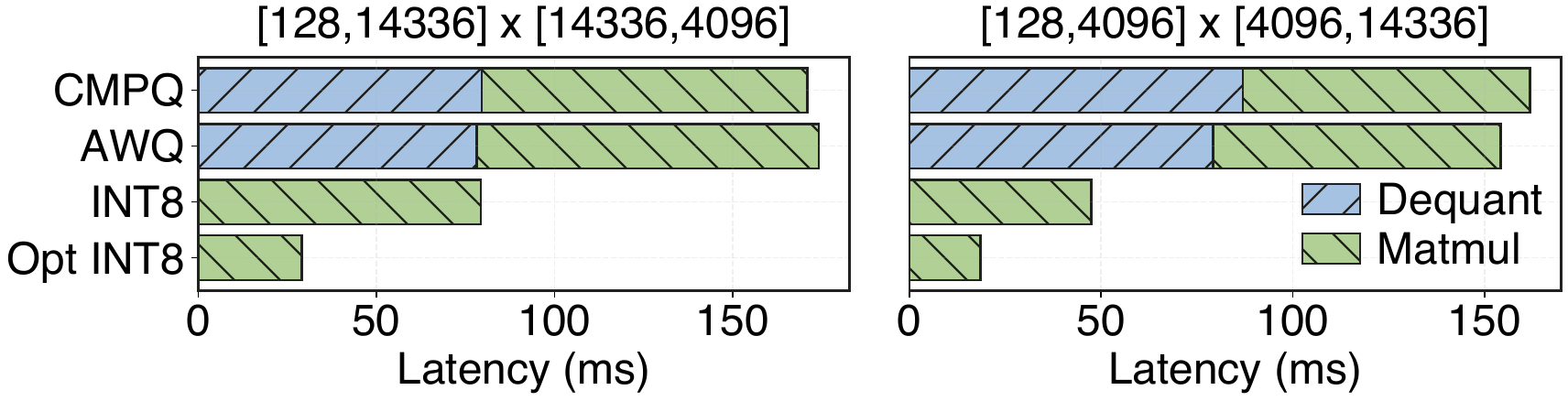}
    \vspace{-8mm}
    \caption{The execution times of NPU matmuls on tensors quantized by AWQ and CMPQ compared with the pure INT8 matmul (\textsf{INT8}) operator and optimized INT8 matmul (\textsf{Opt INT8}) operator.}
    \label{fig:dequant-challenge}
    \vspace{-0.2in}
\end{figure}

We evaluate the execution latency of matmul operators on tensors quantized by AWQ~\cite{mlsys24awq} and CMPQ~\cite{arxiv25cmpq}, two state-of-the-art importance-aware quantization algorithms.
We measure the NPU execution time of matmul on two representative tensor shapes in the inference process of Llama3 8B. 
As shown in Table~\ref{tab:challenge1-constraint}, both approaches only quantize weight tenors. 
AWQ employs \textit{asymmetric per-block} quantization to quantize all weights into INT4.
CMPQ adopts \textit{non-uniform per-channel} quantization on input channels and quantizes weights to the range of 2 to 4 bits with an average 3-bit precision.
However, neither approach can be directly accelerated by the NPU due to the incompatible mapping type and quantization granularity.
When being executed on NPU, quantized weights need to be dynamically dequantized into INT8 with extra arithmetic operations.
As shown in Figure~\ref{fig:dequant-challenge}, AWQ and CMPQ incur $2.59\times$ and $2.63\times$ overhead compared with INT8 matmul on NPUs.
Moreover, during the computation graph optimization phase, the QNN toolchain optimizes the storage layout for static weight tensors to further improve execution efficiency.
The static graph optimization approach cannot be applied to accelerate the two approaches since their tensors must be dynamically dequantized.
As a result, compared with the statically optimized INT8 matmul (\textsf{Opt INT8}), an additional $2.66\times$ slowdown is introduced.

\textbf{Challenge 2: \textit{Costly unpacking for variable-precision weights.}}
NPUs only support a few native data types, \eg INT8 and FP16, which can be executed directly by their tensor accelerators~\cite{qnn}.
When weights are quantized into various precisions, they must first be unpacked into one of these native types, incurring extra computation overhead.

There are two ways to pack and unpack weight tensors, both of which exhibit a trade-off between read amplification and unpacking computation.
First, we can adopt an INT4/INT8 mixed format to pack data by fitting and padding mixed-precision weights into INT4 and INT8 data types. 
The unpacking overhead of this format is minimal since weights are already aligned.
Nevertheless, such an approach incurs substantial read amplifications, as padding causes the system to fetch much more data from flash.

The second approach is adopting the K-Quant~\cite{llama-cpp} format in different weight groups.
K-Quant designs a format for storing single-precision weight tensors with unaligned precisions, \eg INT3.
It decouples the weights into 1-, 2-, and 4-bit components and stores them separately in a compact file.
While such an approach eliminates read amplification, it introduces significant unpacking overhead. 

\begin{figure}[t]
    \footnotesize
    \centering
    \subfloat[The precision distribution for loading and unpacking.]{
        \includegraphics[width=1\columnwidth]{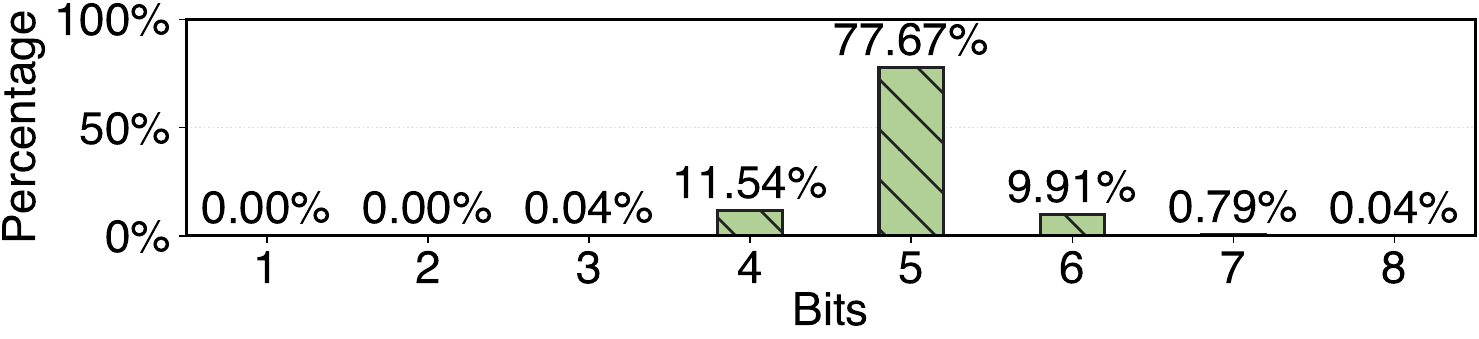}
        \vspace{-2mm}
        \label{fig:unpack-challenge-distribution}
    }
    \vspace{2mm}
    \subfloat[The loading and unpacking overhead.]{
        \includegraphics[width=1\columnwidth]{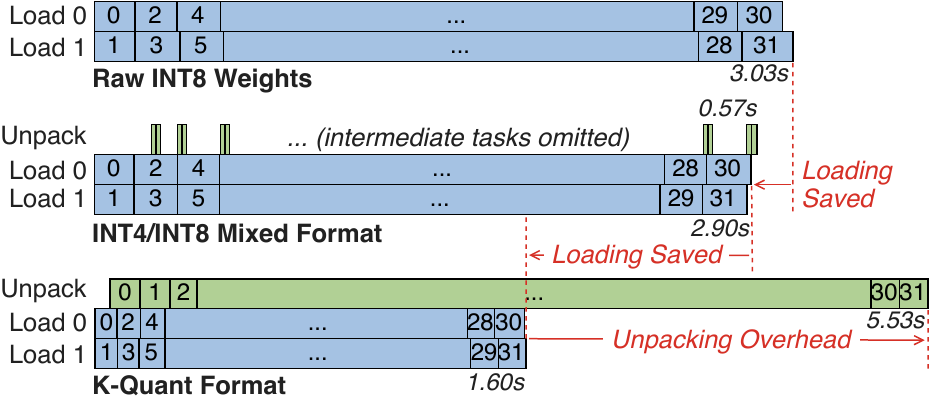}
        \vspace{-1mm}
        \label{fig:unpack-challenge-pipeline}
    }
    \caption{The weight loading and unpacking times of three weight packing schemes for Llama3 8B.}
    \label{fig:unpack-challenge}
    \vspace{-0.2in}
\end{figure}

We measure the overhead of the two approaches for unpacking Llama3 8B quantized by our adaptive precision allocation algorithm.
Figure~\ref{fig:unpack-challenge-distribution} shows the precision distribution of the quantized model, and Figure~\ref{fig:unpack-challenge-pipeline} illustrates their corresponding unpacking overheads.
We use two threads to load the model in a layer-by-layer manner and one thread to unpack the loaded weights.
Each block in the figure represents the loading time or the unpacking time of a single layer.
A number inside a block denotes the layer being loaded or unpacked.
The INT4/INT8 mixed format reduces model loading from 3.03s to 2.90s compared with naively padding all weights into INT8.
K-Quant decreases loading latency to 1.60s due to its compact storage.
However, the unpacking computation takes 5.53s, which becomes a new bottleneck on the critical path of cold starts.

\textbf{Challenge 3: \textit{Imbalance workloads between the CPU and the NPU.}}
Prefill computation constitutes another bottleneck on the critical path of cold starts.
The main challenge is to balance the computation workload between the CPU and NPU while keeping minimal idle time.
Nevertheless, due to the heterogeneous nature of operators during the LLM inference computation and their varying computational characteristics, achieving such a balance is challenging.

\begin{figure}[t]
    \centering
    \begin{subfigure}{0.98\columnwidth}
        \centering
        \includegraphics[width=\linewidth]{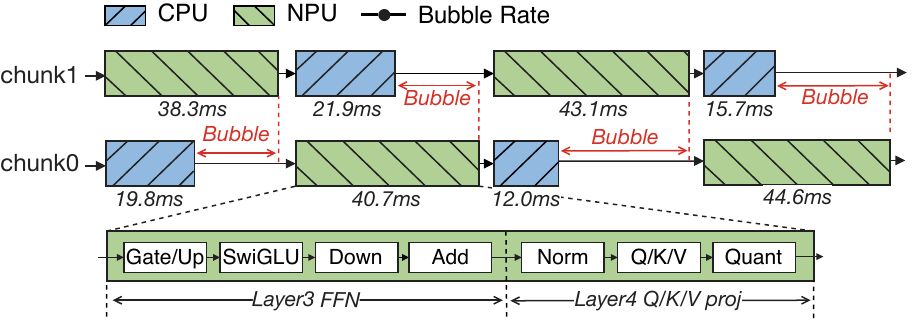}
        \caption{A fragment of the measured pipeline of llm.npu.}
        \label{fig:mllm-challenge:a}
    \end{subfigure}
    \vspace{0.5em}
    \begin{subfigure}{0.38\columnwidth}
        \centering
        \includegraphics[width=\linewidth]{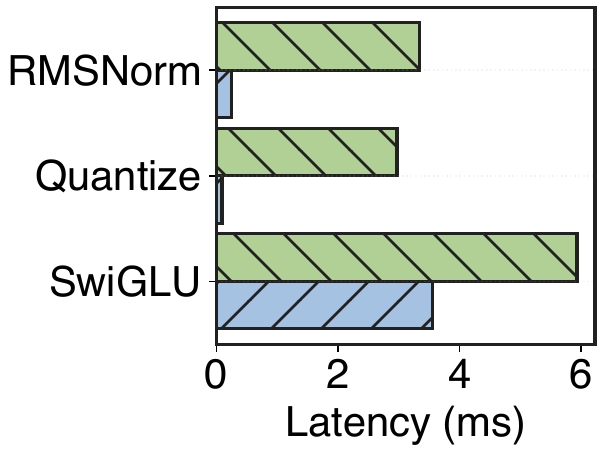}
        \caption{Execution times of NPU-inefficient operators on the NPU and CPU, respectively.}
        \label{fig:mllm-challenge:b}
    \end{subfigure}
    \hfill
    \begin{subfigure}{0.58\columnwidth}
        \centering
        \includegraphics[width=\linewidth]{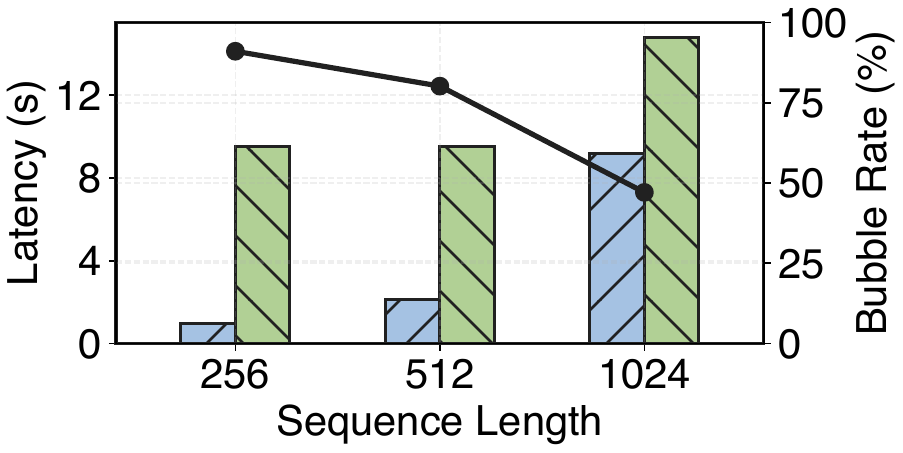}
        \caption{The prefill computation time of the NPU and CPU, respectively, and the bubble rate of the entire pipeline.}
        \label{fig:mllm-challenge:c}
    \end{subfigure}
    \vspace{-0.1in}
    \caption{Analyses of the pipeline of llm.npu during the prefill stage of Llama3 8B.}
    \label{fig:mllm-challenge}
    \vspace{-0.25in}
\end{figure}

The state-of-the-art approach, \ie llm.npu~\cite{asplos25mllm}, partitions the LLM computation graph in a coarse-grained and static manner among the NPU and the CPU.
We evaluate llm.npu with a 256-token prefill on Llama3 8B for all subgraphs and measure its CPU and NPU computation time, as shown in Figure~\ref{fig:mllm-challenge}.
llm.npu adopts chunked prefill~\cite{osdi24sarathi-serve} and parallelizes the computation of two adjacent chunks on the CPU and NPU in the granularity of statically partitioned subgraphs.
Only attention computations are placed on the CPU side, while the remaining operators between layers run on the NPU.
As shown in Figure~\ref{fig:mllm-challenge:a}, such a design leaves substantial pipeline bubbles for two main reasons.
First, coarse-grained partitioning assigns many NPU-inefficient operators to the NPU, \ie operators with low arithmetic intensity, which causes high NPU execution time.
Figure~\ref{fig:mllm-challenge:b} shows the CPU and NPU execution time of three representative operators that llm.npu assigns to the NPU, \ie RMSNorm, SwiGLU, and quantization.
The NPU execution time is on average $2.1\times$ longer than that on the CPU.
Second, static scheduling prevents dynamic adjustment based on runtime status.
Figure~\ref{fig:mllm-challenge:c} illustrates the CPU and NPU execution time and the bubble rate of the entire pipeline across different sequence lengths.
The bubble rate is calculated as the proportion of idle time to the execution time.
The NPU execution time is consistently longer than the CPU execution time, resulting in up to 91\% bubble rates across the pipeline.
Moreover, this imbalance varies with the prompt length, making static scheduling incapable of maintaining balanced utilization.

\section{The \EdgeFlow Design}\label{sec:design}

\begin{figure}[!t]
    \centering
    \includegraphics[width=1\columnwidth]{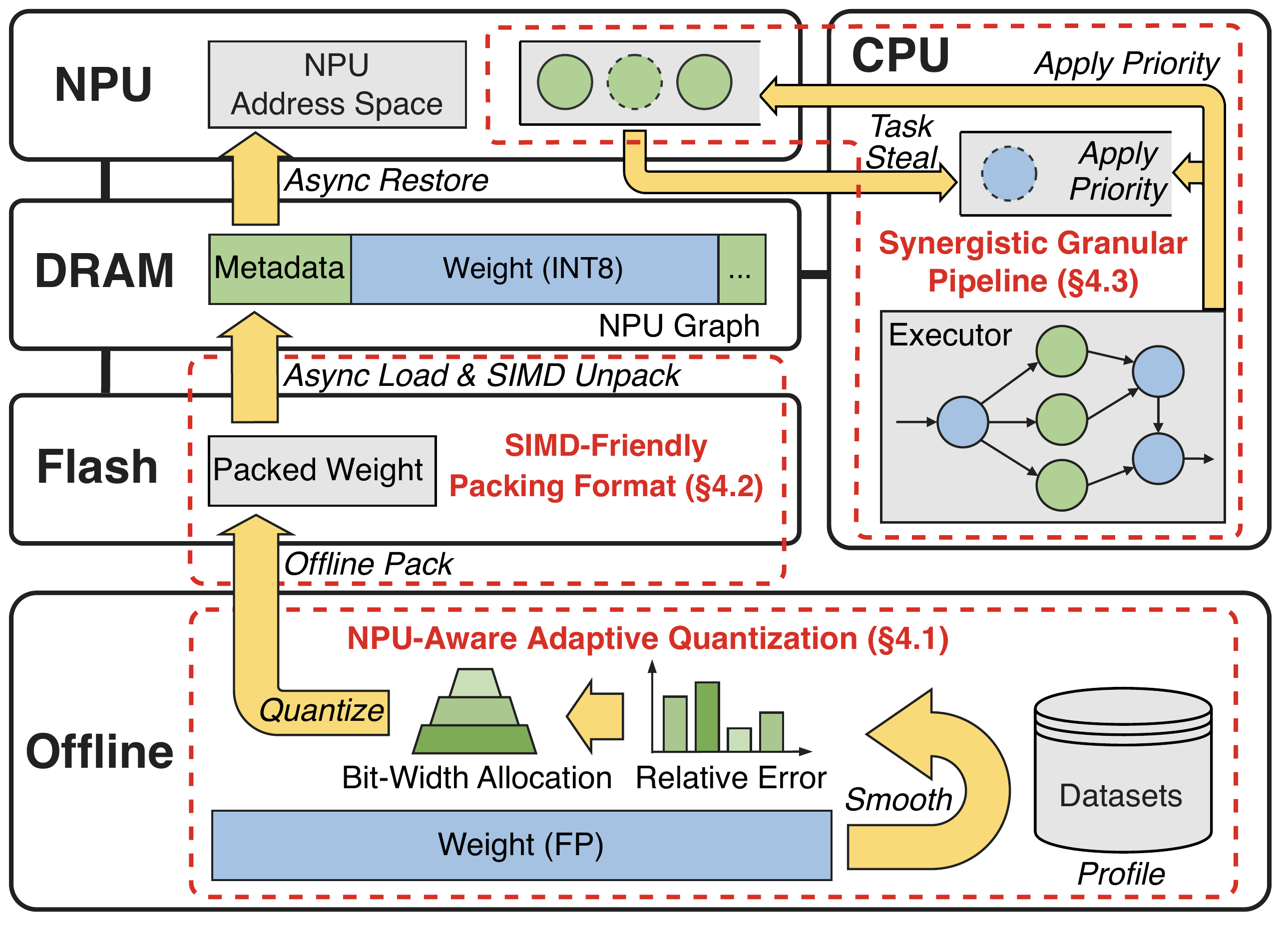}
    \caption{The overview of \EdgeFlow.}
    \label{fig:overview}
    \vspace{-0.15in}
\end{figure}

\noindent
We design \EdgeFlow to address all of the above challenges and reduce the cold-start latency of mobile LLM inferences.
As illustrated in Figure~\ref{fig:overview}, \EdgeFlow consists of two phases, \ie an offline phase and an online phase.
The offline phase performs weight quantization to accelerate model loading on the critical path of cold starts.
We introduce an \textit{NPU-aware adaptive quantization} method to achieve a better trade-off between the loading time and the model accuracy under NPU constraints (\textbf{Challenge 1}).

The online phase is responsible for preparing the execution graph and conducting the prefill computation.
To accelerate the process, data loading, weight unpacking, and prefill computation are executed in parallel.
The quantized model is loaded asynchronously from flash and unpacked to NPU-native formats layer by layer.
An \textit{SIMD-friendly packing format} is proposed to store weights, and an SIMD-based unpacking algorithm is adopted to accelerate the unpacking process (\textbf{Challenge 2}).
Moreover, a \textit{synergistic granular pipeline} is employed during prefill computation to balance workloads between the CPU and the NPU and alleviate the computation bottleneck of cold starts (\textbf{Challenge 3}).

\subsection{NPU-Aware Adaptive Quantization}
\label{sec:rime}

\noindent
As mentioned in Section~\ref{sec:challenge}, NPUs only accelerate matmuls for per-output-channel quantized weight tensors and coarse-grained per-tensor quantized activation tensors.
Our quantization scheme complies with these restrictions and achieves a better trade-off between model size and accuracy with two design choices:
1) We assign different precisions in the granularity of output channels on weight tensors to satisfy the restrictions of NPUs on weights.
Each output channel is quantized into integers with different bit-widths.
A \textit{relative error metric} is defined to guide the precision assignment on each channel.
In addition, a \textit{greedy bit-width allocation} algorithm is proposed to derive an optimal assignment of precisions across all channels.
2) We transfer the quantization difficulty of both input and output activations to weights to respect the NPUs' restrictions on activation tensors.
An \textit{NPU-aware smoothing} scheme is adopted to achieve this goal without additional runtime overhead.

\noindent\textbf{The Relative Error Metric.}
We design a relative error metric to estimate the amount of quantization error a weight channel introduces when quantized to a specific bit-width.
Channels with higher metric values are granted more bits.
Formally, for channel $i$ with $D$ dimensions, we define the relative error under $B$-bit quantization, \ie $\text{RE}(W_i, B)$, as the cosine distance between the original weight channel $W_i$ and its dequantized version $W^q_i$~\cite{nips188bit-training, nips19tnt}:
\setlength{\abovedisplayskip}{5pt}
\setlength{\belowdisplayskip}{5pt}
\[
\text{RE}(W_i,B) = 1 - \frac{W_i \cdot W^q_i}{\|W_i\| \|W^q_i\|}.
\]

However, computing the cosine distance for all channels under all candidate bit-widths is expensive, which results in high quantization overheads.
Particularly, before computing the cosine distance, we first need to get $W^q_i$ by uniformly quantizing $W_i$ into $B$ bits and dequantizing it back to FP16.
Two linear transformations and one cosine distance computation are involved in the process of calculating $\text{RE}$ for one channel under a single bit-width. 
We reduce the computation overhead by modeling $\text{RE}$ with random variables so that it can be efficiently estimated with simple statistics over $W_i$ and $B$ instead of calculating the precise values.

To achieve this, we approximate $\text{RE}$ with the absolute error $E = W_i - W^q_i$ and model $E$ as a random variable.
We skip the detailed derivation in Appendix~\ref{appendix:relative_error_derivation}.
The $\text{RE}$ can be represented as a function of the mathematical expectations of $E^2$ and ${W_i}^2$:
\setlength{\abovedisplayskip}{5pt}
\setlength{\belowdisplayskip}{5pt}
\[
\text{RE}(W_i,B) = 1 - \frac{W_i \cdot (W_i-E)}{\|W_i\| \cdot \|W_i-E\|} \approx \frac{\mathbb{E}[E^2]}{2 \, \mathbb{E}[{W_i}^2]}.
\]

We further adopt a uniform rounding approximation~\cite{mlsys24awq} to simplify the calculation of $\mathbb{E}[E^2]$.
Formally, we model $E_j$, \ie the absolute error of the $j$-th element in $W_i$, with $E_j \sim \text{U}(\frac{-S_i}{2}, \frac{S_i}{2})$, where $S_i$ is the scale of channel $i$ under uniform quantization.
The rationale is that the absolute error always falls in the range $[\frac{-S_i}{2}, \frac{S_i}{2}]$ and $S_i$ is usually small.
We can use a uniform distribution to approximate the real distribution in a small range.
By applying the uniform estimation to $\mathbb{E}[E^2]$, we obtain the final form:
\setlength{\abovedisplayskip}{5pt}
\setlength{\belowdisplayskip}{5pt}
\[
\text{RE}(W_i, B) = \frac{1}{2^{2B}} \cdot \frac{(\max|W_i|)^2}{\mathbb{E}[{W_i}^2]}.
\]
The advantage of this expression is that it can be calculated with simple statistics, \ie mean squared values, which greatly simplifies the computation.

\noindent\textbf{Greedy Bit-Width Allocation.}
For a weight tensor with $C$ output channels $\{W_i\}_{i=1}^C = \{W_1, W_2, ..., W_c\}$, we need to find the optimal per-channel bit-widths $\{B_i\}_{i=1}^C = \{B_1, B_2, ..., B_c\}$ that minimizes the total $\text{RE}$.
Two constraints need to be satisfied.
First, the assignable bit-widths for each channel should be in the range 1 to 8 bits. 
Second, the average bit-width of all channels should not exceed a given value $B_e$ to indicate the flash bandwidth budget.
We formulate the optimization problem as follows:
\setlength{\abovedisplayskip}{5pt}
\setlength{\belowdisplayskip}{5pt}
\[
\begin{aligned}
\min_{\{B_i\}_{i=1}^C} \quad & \sum_{i=1}^C \text{RE}(W_{i},B_i), \\
\text{s.t.} \quad & \sum_{i=1}^C B_i \le C \cdot B_e, \; B_i \in \{1, 2, \dots, 8\}.
\end{aligned}
\]

We leverage two properties of this optimization problem to design our bit-width allocation algorithm. 
First, the total $\text{RE}$ in our formulation is additive across channels, meaning the $\text{RE}$ of one channel is determined solely by its own bit-width and statistics, independent of others. 
While end-to-end model accuracy is indeed a complex combinatorial function of all quantizations, we use this additive $\text{RE}$ as a local proxy metric to make the optimization tractable.
Second, the reduction in $\text{RE}$ for a channel is strictly decreasing with more bit-width $B$ assigned to it. 
These two properties guarantee that we can greedily assign a bit to one channel that reduces the total $\text{RE}$ most without sacrificing the global optimality.

Therefore, we develop a greedy algorithm that yields the optimal solution.
As shown in Algorithm~\ref{alg:bit_allocation}, we first initialize all channels to 1 bit and then compute the remaining bit budget (Line~1).
We build a max heap using the $\text{RE}$ gain of adjacent bit-widths for each channel as key (Line~2), \ie $\text{RE}(W_i, B) - \text{RE}(W_i, B+1)$.
We then iteratively pop the channel that reduces its relative error the most from the heap, increment its bit-width by one, update the heap, and adjust the remaining bit-width budget (Lines~3-7).
The process continues until the total bit-width budget is exhausted.
This approach efficiently finds an optimal bit-width allocation for $C$ weight channels in $O\big(C\cdot \log C\big)$ time.

\definecolor{darkgreen}{RGB}{0, 102, 5}
\newcommand{\myCmSty}{\color{darkgreen}\itshape\small}
\SetCommentSty{myCmSty}

\SetKwInOut{KwIn}{Input}
\SetKwInOut{KwOut}{Output}

\DontPrintSemicolon
\SetVlineSkip{0.1em}

\begin{algorithm}[t]
    \small
    \KwIn{
        $\{W_i\}^C_{i=1}$: The weight with $C$ channels \\
        $budget$: The expected average bit-width
    }
    \KwOut{
        $\{B_i\}_{i=1}^{C}$: The allocated bit-width array
    }

    \tcp{Initialize all channels to 1 bit and calculate expected gains}
    $FillOnes(B),\ remain \leftarrow budget - 1$\;
    $\mathcal{H} \leftarrow MaxHeap(\{\, \mathrm{RE}(W_i,1) - \mathrm{RE}(W_i,2)\}^C_{i=1})$\;
    \While{$remain > 0$}{
        \tcp{Greedily allocate 1 more bit to the max-gain channel $j$}
        $gain_j \leftarrow \mathcal{H}.pop()$\;
        $B_j \leftarrow B_j + 1,\ remain \leftarrow remain - \frac{1}{C}$\;
        \If(\tcp*[f]{Update the gain of channel $j$}){$B_j < 8$}{
            $\mathcal{H}.push(\text{RE}(W_j, B_j) - \text{RE}(W_j, B_j+1))$\;
        }
    }
    \Return $B$\;

\caption{Greedy Bit-Width Allocation}
\label{alg:bit_allocation}
\end{algorithm}

\noindent\textbf{NPU-Aware Smoothing.}
NPUs' coarse-grained per-tensor quantization degrades accuracy severely on high-variance LLM activations~\cite{asplos25mllm,asplos25comet}.
Meanwhile, our bit-width allocation algorithm on weights is input-unaware, making it insufficient to address activation-induced errors.
To tackle these issues, we propose NPU-aware smoothing, which statically transfers inter-channel variance from activations to weights without runtime overhead.

We use a calibration dataset to profile per-channel variance $S_I\in \mathbb{R}^{D}$ and $S_O\in\mathbb{R}^{C}$ for the input $I$ and output $O$, respectively.
The variance for each channel is defined as the maximum absolute value in the channel.
We then divide inputs and outputs by ${S_I}^{\alpha}$ and ${S_O}^{\beta}$ so that their variances are smoothed. 
$\alpha$ and $\beta$ are hyperparameters controlling the degree of smoothing.
The variances are then restored in a transformed weight tensor $W' = \mathrm{diag}({S_I}^\alpha) \cdot W \cdot \mathrm{diag}({S_O}^{-\beta})$.
The resulting quantized matmul becomes:
$$
O = (I \cdot \mathrm{diag}({S_I}^{-\alpha})) \cdot W' \cdot \mathrm{diag}({S_O}^\beta)
$$
Performing bit-width allocation on the restored weight tensor $W'$ is implicitly guided by activation statistics, making the overall scheme input-aware, similar in spirit to AWQ~\cite{mlsys24awq}.

We perform a grid search to find the optimal $\alpha$ over the interval of $[0,\,1] \in \mathbb{R}$ that minimizes quantization error.
We set $\beta$ to 1 to shift most of the inter-channel variance of outputs to weights.
Moreover, the scaling factors do not introduce additional runtime overhead, as the input-side scaling can be fused into the preceding normalization or linear operators, while the output-side scaling can be absorbed by the subsequent dequantization step.

\subsection{SIMD-Friendly Packing Format}
\label{sec:format}

\begin{figure}[t]
    \centering
    \includegraphics[width=1\columnwidth]{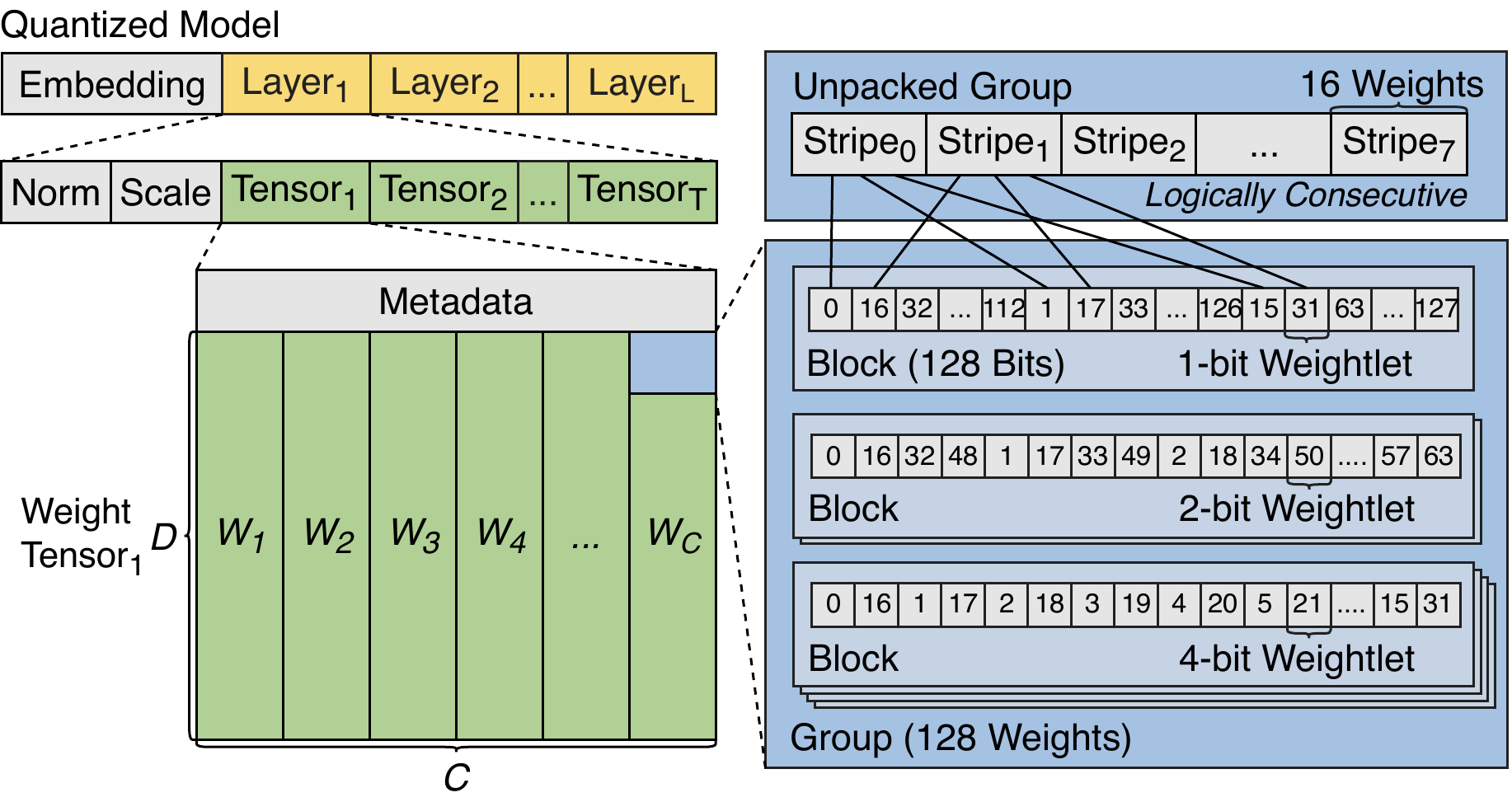}
    \vspace{-7mm}
    \caption{The SIMD-friendly packing format. The number denoted in each weightlet is the index of its corresponding weight.}
    \label{fig:packing-format}
    \vspace{-0.25in}
\end{figure}

The unpacking process converts low-bit weights into INT8 tensors for NPU execution.
\EdgeFlow accelerates this process with SIMD-based unpacking.
SIMD (Single Instruction Multiple Data) provides wide registers, \eg 128-bit registers in Arm Neon~\cite{arm-neon}, to apply the same operation to multiple elements in parallel.

However, SIMD unpacking faces two challenges.
First, irregular bit-widths such as 3-bit or 5-bit do not align naturally with byte boundaries.
We address this issue by decomposing each weight into a combination of primitive \textit{weightlets} with bit-widths 1, 2, or 4, so that unpacking reduces to handling only byte-divisible primitives.
Second, SIMD manipulates data at byte granularity, which makes it inefficient to directly access logically adjacent weightlets stored contiguously.
We therefore store weightlets in an interleaved layout.
For an $R$-bit SIMD register, this layout allows processing up to $R/8$ weightlets in parallel and reconstructing $R/8$ consecutive weights simultaneously.

Figure~\ref{fig:packing-format} illustrates the resulting storage format of a quantized model that contains $L$ layers, each with multiple weight tensors.
Each tensor is a $D \times C$ matrix, where $C$ is the number of output channels and $D$ is the number of weights per channel.
Within each channel, weights are partitioned into \textit{groups} of $R$ consecutive weights.
After decomposition, weightlets of the same bit-width in a group are stored into one or more $R$-bit \textit{blocks}, each directly loadable into an SIMD register.
For example, a group of 7-bit weights is decomposed into four 4-bit blocks, two 2-bit blocks, and one 1-bit block.
To match SIMD byte-wise execution, weightlets belonging to $R/8$ logically consecutive weights (defined as a \textit{stripe}), are interleaved with a stride $8/B$ for bit-width $B \in \{1,2,4\}$.
Since different channels may use different bit-widths, each packed tensor is prepended with per-channel bit-width metadata, stored as a compact INT3 array representing eight precision levels from 1 to 8 bits.

\begin{figure}[t]
    \centering
    \includegraphics[width=1\columnwidth]{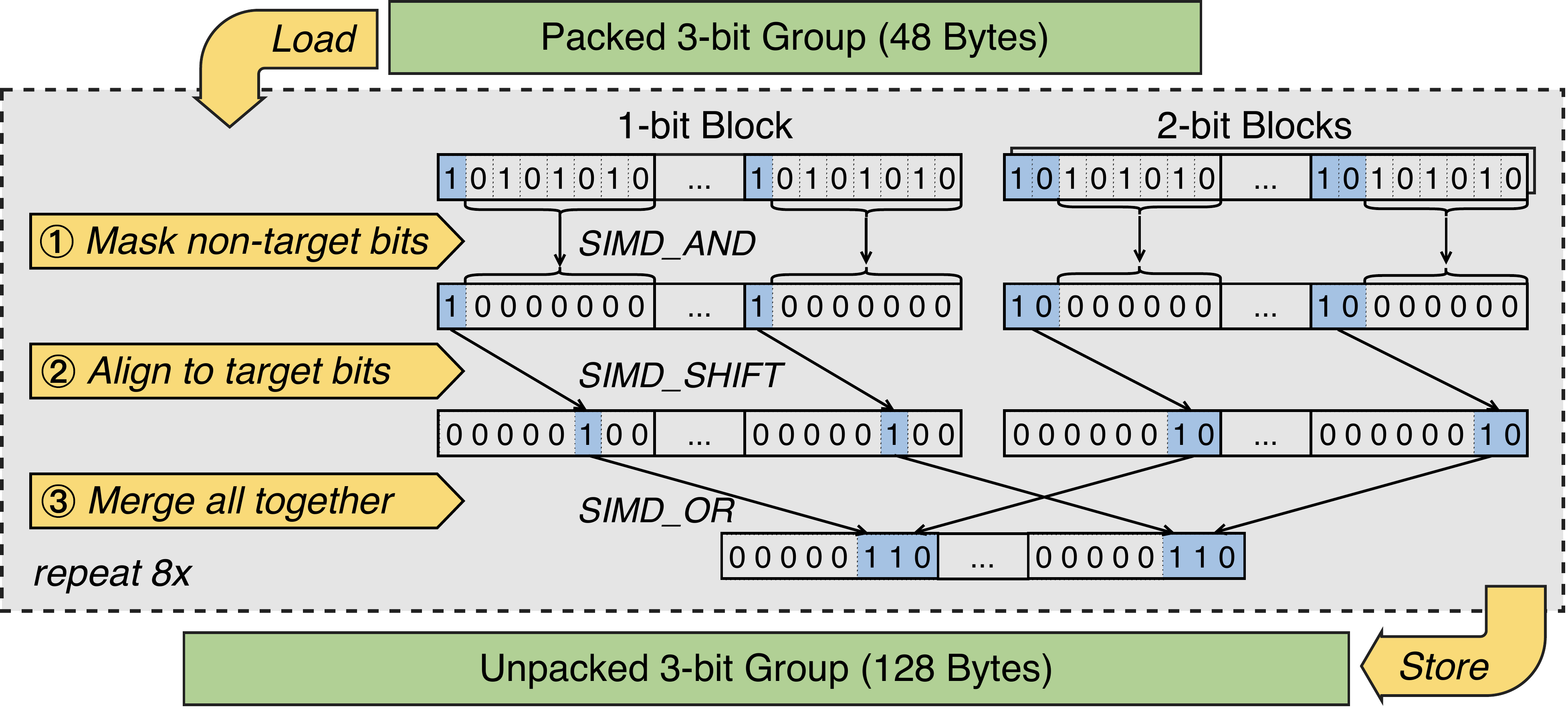}
    \vspace{-5mm}
    \caption{The SIMD-based unpacking algorithm.}
    \label{fig:unpack-algo}
    \vspace{-0.25in}
\end{figure}

\EdgeFlow further designs an SIMD-based unpacking algorithm for efficient weight unpacking.
The process is also conducted in a channel-wise, group-wise manner.
First, \EdgeFlow checks the metadata of the current channel to determine its bit-width, which indicates the number of blocks per group.
Figure~\ref{fig:unpack-algo} illustrates the unpacking procedure exemplified by a 3-bit weight group, which consists of one 1-bit block and two 2-bit blocks, each mapped to a wide register.

For each stripe, unpacking proceeds in three steps.
Taking the first stripe as an example:
\textcircled{\footnotesize 1}~Mask. Valid bits are extracted via bitwise \textit{AND}s: the leading bit of each byte in the 1-bit block and the leading two bits in the first 2-bit block.
\textcircled{\footnotesize 2}~Align. The extracted bits are shifted to their target positions: the 1-bit block is right-shifted by 5 bits to form the MSB, while the 2-bit block is right-shifted by 6 bits to form the lower bits.
\textcircled{\footnotesize 3}~Merge. The aligned blocks are combined via bitwise \textit{OR}s to form 3-bit weights, then cast to INT8.

This process is repeated across all stripes (\eg 8$\times$ per group) until the entire group is unpacked and stored for NPU execution.
With this SIMD-based design, unpacking requires only 0.48 SIMD instructions per weight on average.

\subsection{Synergistic Granular Pipeline}
\label{sec:scheduler}
\begin{figure*}[t]
    \centering
    \begin{tikzpicture}
        \node[anchor=south west, inner sep=0] (img)
            {\includegraphics[width=\textwidth]{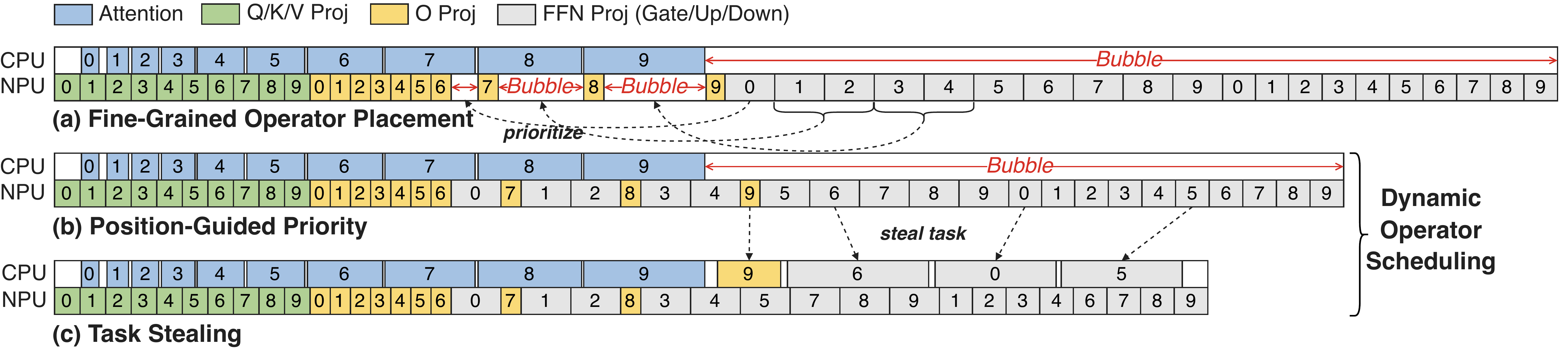}};
        \phantomsubcaption
        \label{fig:schedule:a}
        \phantomsubcaption
        \label{fig:schedule:b}
        \phantomsubcaption
        \label{fig:schedule:c}
    \end{tikzpicture}
    \vspace{-5mm}
    \caption{The synergistic granular pipeline with fine-grained operator placement and dynamic operator scheduling. Each block represents an operator, and the numbers inside indicate chunk IDs.}
    \label{fig:schedule}
    \vspace{-0.1in}
\end{figure*}

\noindent
As mentioned in Section~\ref{sec:challenge}, the key problems with existing pipeline scheduling approaches originate from their coarse-grained and static nature.
\EdgeFlow attacks the issue by first designing a fine-grained operator placement scheme that enables operator scheduling with more flexibility.
Then, a dynamic operator scheduling mechanism is adopted to further reduce the pipeline bubbles.

\noindent\textbf{Fine-Grained Operator Placement.}
\EdgeFlow performs scheduling at the granularity of individual operators, \eg matmuls and layer normalizations.
Compared with scheduling at subgraph granularity, such an approach enables finer-grained control over computation.
In addition, we derive an initial placement of all operators among the NPU and the CPU to serve as the static backbone for the subsequent dynamic scheduling.
We place all INT8 matmul operators on the NPU, \ie Q/K/V/O projections for attentions and Gate/Up/Down projections for feedforward networks (FFNs), since NPUs are efficient in executing integer matmuls.
Other operators are assigned to the CPU in FP16.
To respect operator dependencies, we offline order operators topologically within a transformer layer.
We maintain separate operator queues for the CPU and the NPU, and dispatch operators according to their topological order during runtime.
Detailed placement is provided in Appendix~\ref{appendix:operator_level_placement}.

\noindent\textbf{Dynamic Operator Scheduling.}
To further optimize resource utilization, \EdgeFlow proposes to dynamically schedule operators between the CPU and the NPU.
Specifically, we adopt a position-guided operator priority to reduce NPU bubbles and a task-stealing scheme to reduce CPU bubbles.

The position-guided operator priority reduces NPU bubbles caused by the suboptimal operator execution order.
Specifically, existing approaches serialize the operator execution of input sequence chunks and parallelize chunk execution on the CPU and NPU.
As shown in Figure~\ref{fig:schedule:a}, the NPU first executes the Q/K/V projections for all chunks.
Once the Q/K/V tensors are produced for a chunk, the CPU then computes the attention of the chunk in parallel.
The O projection is then executed in parallel on the NPU when the corresponding attention score is ready for the chunk.
Since the computation time of attentions grows linearly with the number of chunks, O projections on NPUs cannot overlap with attention computations on the CPU as the chunk count increases.
NPU time is thus wasted on waiting for the CPU attention computation, even though the subsequent FFN projection of previous chunks can be executed first.

The root cause of this issue lies in the lack of a decent tie-breaking mechanism when scheduling operators under the topological order of models' execution graphs, \eg the case when inputs of both O projection for later chunks and FNN projections for earlier chunks are ready.
\EdgeFlow breaks the tie by assigning operators a positional-guided priority.
As shown in Figure~\ref{fig:schedule:b}, for each operator, we assign a priority based on its chunk position, where earlier chunks in the prompt receive higher priority.
This strategy effectively breaks the tie by executing CPU operators in earlier chunks and helps unlock downstream NPU operators earlier.

Moreover, we find that the CPU remains idle for 65\% of the end-to-end pipeline execution time.
We thus implement a CPU task-stealing mechanism to enhance CPU utilization.
As shown in Figure~\ref{fig:schedule:c}, when the CPU is idle and the length of the NPU task queue exceeds a threshold, the CPU proactively takes and executes the top task from the NPU task queue.
The threshold is adopted to avoid excessive CPU preemption that may leave the NPU idle.
We determine the threshold according to the ratio between the execution time of CPU and NPU matmul operators and set it to 5 in our experiments.
\section{Evaluation}\label{sec:evaluation}

\subsection{Experimental Setup}
\label{sec:experimental_setup}

\noindent\textbf{Implementations.}
We implement \EdgeFlow from scratch with about 8K lines of code. 
\EdgeFlow is composed of 1) a set of Python utilities that quantize models and pack weights in the SIMD-friendly packing format, and 2) a C++ runtime that unpacks weights and conducts inference computation.
The current runtime is implemented with QNN and supports mobile devices with Arm CPUs and Hexagon NPUs.
QNN embeds weights into compiled execution graphs in a proprietary storage format during the graph compilation phase. 
There is no interface for dynamically loading weights into a compiled execution graph.
However, \EdgeFlow stores quantized weights in an SIMD-friendly unpacking format.
Weights need to be dynamically unpacked into execution graphs before computation.
We reverse-engineer the proprietary execution graph format of QNN to enable dynamic loading and execution of weights by the NPU.

\noindent\textbf{Testbed.}
All experiments are conducted on a Xiaomi 15 Pro~\cite{xiaomi15pro} equipped with a Qualcomm Snapdragon 8 Elite SoC, 16 GB of RAM, and 512 GB of flash storage.
Inside the Snapdragon SoC, there is an Arm CPU with 8 cores, a Hexagon NPU, and an Adreno GPU.

\begin{table}[t]
\centering
\footnotesize
\setlength{\tabcolsep}{2pt}
\caption{Datasets used in our evaluation. "OBQA" is OpenBookQA.}
\label{tab:eval-datasets}
\vspace{-3mm}
\begin{tabular}{lccccc}
\toprule
Datasets & LAMBADA & WinoGrande & OBQA & MMLU & HellaSwag \\
\midrule
\#shots & -- & 3 & 4 & 5 & 6 \\
\#token range & 59--97 & 131--144 & 190--230 & 243--699 & 883--1086 \\
\bottomrule
\vspace{-0.25in}
\end{tabular}
\end{table}

\noindent\textbf{Models and Datasets.}
We evaluate \EdgeFlow under models with sizes that are widely used on mobile devices, \ie Llama3 8B~\cite{Llama3-8B}, Mistral 7B~\cite{Mistral-7B}, Phi3 3.8B~\cite{Phi3-3.8B}, and Qwen1.5 1.8B~\cite{Qwen1.5-1.8B}.
As shown in Table~\ref{tab:eval-datasets}, we utilize five datasets, \ie LAMBADA~\cite{acl16lambada}, WinoGrande~\cite{aaai20windogrande}, OpenBookQA~\cite{acl18obqa}, MMLU~\cite{iclr21mmlu}, and HellaSwag~\cite{acl19hellaswag}, like prior works~\cite{asplos25mllm,eurosys26llamanpu,iclr23gptq}, to evaluate the model accuracy and the cold start latency of \EdgeFlow under various prompt lengths.
We adopt multi-shot prompting to generate prompt lengths in a more flexible manner.
The summarized prompt lengths are shown in Table~\ref{tab:eval-datasets}.

\noindent\textbf{Baselines.}
We evaluate \EdgeFlow against llama.cpp~\cite{llama-cpp}, MNN~\cite{mmasia24mnn}, and llm.npu~\cite{asplos25mllm}.
llama.cpp and MNN are popular open-source mobile LLM frameworks.
llama.cpp maps model weights into memory with mmap and loads weights with on-demand paging during cold starts.
MNN loads weights sequentially before the inference computation.
Due to the bandwidth-bound nature during cold starts and the lack of NPU support in these two frameworks, we use the more widely adopted CPU backend for both baselines.
llm.npu is the state-of-the-art NPU-based mobile LLM framework.
It constructs and optimizes computation graphs on the critical paths of cold starts, which incur substantial graph compilation overhead.
We augment llm.npu with materialization and overlapping as mentioned in Section~\ref{sec:challenge} for more fair comparisons.

\begin{figure*}[!t]
    \centering
    \includegraphics[width=\textwidth]{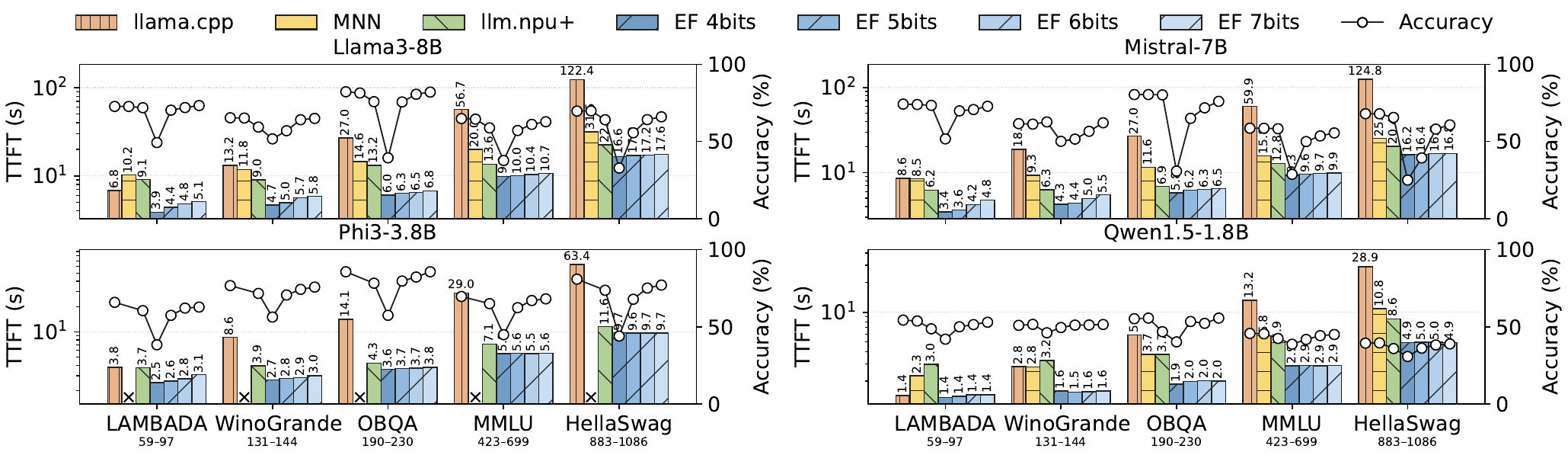}
    \vspace{-7mm}
    \caption{The cold start latency (\ie TTFT) and accuracy of different methods on various models and datasets. Bars show the average TTFT (the lower is better). Lines show accuracy. Prompt length is shown beneath each dataset label. "EF" represents \EdgeFlow. "llm.npu+" represents llm.npu enhanced with materialization and overlapping techniques; the same notation is used throughout.}
    \label{fig:main_performance}
    \vspace{-0.15in}
\end{figure*}

\subsection{End-to-End Performance}\label{sec:performance}

\noindent
Figure \ref{fig:main_performance} presents the cold start latency (TTFT) of \EdgeFlow and all baseline systems under different models and datasets.
For \EdgeFlow, we present both TTFT and model accuracy when the model is quantized to an average of 4 to 7 bits and discuss the results from two perspectives.

\noindent\textbf{The Cold-Start Latency.}
As the bit-width decreases from 7 bits to 4 bits, \EdgeFlow progressively achieves higher speedups.
Specifically, at 7 bits, it reduces TTFT by $3.92\times$, $2.28\times$, and $1.47\times$ compared with llama.cpp, MNN, and llm.npu, respectively.
While at 4 bits, the speedups increase to $4.24\times$, $2.53\times$, and $1.63\times$.
In datasets with short prompts, \ie under LAMBADA and WinoGrande datasets, TTFT is reduced since flash bandwidth is utilized more efficiently.
This improvement comes from adaptive quantization, which maintains comparable model accuracy while minimizing the amount of data being loaded.
When prompts are long, \ie under MMLU and HellaSwag datasets, the prefill computation becomes the bottleneck.
The TTFT of \EdgeFlow outperforms llm.npu with speedups ranging from $1.37\times$ to $1.41\times$ due to our fine-grained and dynamic pipeline design.

\noindent\textbf{The Model Accuracy.}
All baselines adopt INT8 quantization on model weights, which results in larger model sizes than \EdgeFlow.
Specifically, llama.cpp adopts per-block quantization only on weights.
MNN applies static per-channel quantization on weights and dynamic per-channel quantization on activations.
Finally, llm.npu adopts per-tensor INT8 quantization with a shadow outlier scheme.
Specifically, it first quantizes weights and activations uniformly in the per-tensor granularity and executes the quantized matmuls on the NPU.
For activations with abnormal magnitudes, \ie outliers, llm.npu executes them in FP16 on the CPU.

As shown in the figure, when quantized to an average of 4 to 7 bits, \EdgeFlow gets an accuracy drop of 22.48\%, 6.36\%, 1.91\%, and -0.03\%, respectively, compared with three baselines.
We use 5-bit as the default configuration of \EdgeFlow since it delivers substantial TTFT improvements under comparable accuracy with the NPU-based baseline, \ie llm.npu.

\subsection{Ablation Study}\label{sec:ablation_study}

\begin{figure*}[!t]
    \centering
    \includegraphics[width=\textwidth]{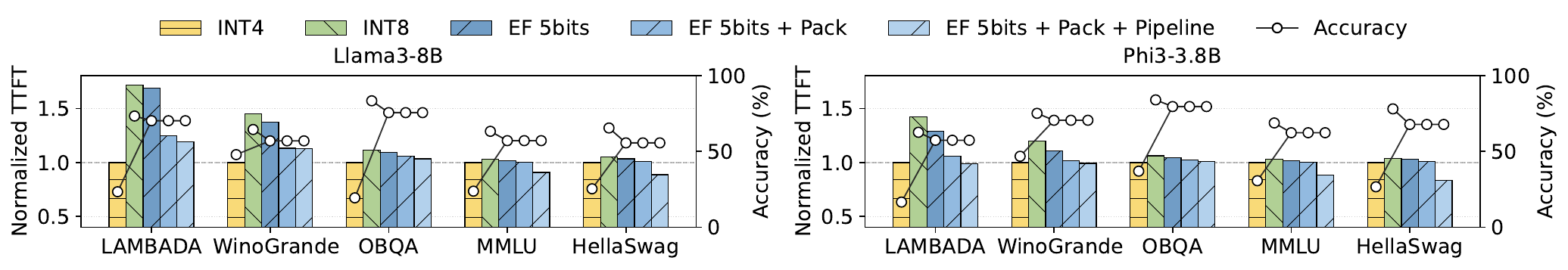}
    \vspace{-7mm}
    \caption{The model accuracy and the normalized TTFT of INT4, INT8, and different techniques in \EdgeFlow.}
    \label{fig:ablation_study}
    \vspace{-0.15in}
\end{figure*}

\noindent
We conduct a comprehensive ablation study to evaluate the contributions of each technique to \EdgeFlow's overall performance with Llama3 8B and Phi3 3.8B over all datasets.
The results are shown in Figure~\ref{fig:ablation_study}.
We use INT8 and INT4 per-tensor quantization with NPU-aware smoothing as our baselines, \ie \textsf{INT8} and \textsf{INT4} in the figure.
\textsf{EF 5bits} denotes \EdgeFlow with only NPU-aware adaptive quantization (Section~\ref{sec:rime}) with an average bit-width of 5 bits.
The quantized weights are stored with the INT4/INT8 mixed packing format.
\textsf{Pack} and \textsf{Pipeline} indicate the incorporation of the SIMD-friendly packing format (Section~\ref{sec:format}) and the synergistic granular pipeline (Section~\ref{sec:scheduler}), respectively.
We observe that each design component contributes significantly to reducing TTFT or maintaining high accuracy, as detailed below.

\noindent\textbf{+ NPU-Aware Adaptive Quantization.}
Compared with \textsf{INT4}, the NPU-aware adaptive quantization scheme improves model accuracy by an average of 35.6\%.
Compared with \textsf{INT8}, model accuracy only drops by 6.6\% on average.
These results show the effectiveness of the NPU-aware adaptive quantization scheme.
Specifically, with just a 1-bit increase in average bit-width, model accuracy can be significantly enhanced.
However, the reduction in TTFT compared with \textsf{INT8} is only 3.3\% due to the lack of an efficient packing format.
Over 88\% of weights are padded to 8 bits, leading to severe read amplifications during model loading.

\noindent\textbf{+ SIMD-Friendly Packing Format.}
After adopting our packing format, TTFT decreases by up to $1.36\times$ and $1.22\times$ compared with \textsf{EF 5bits} under Llama3 8B and Phi3 3.8B, respectively.
This effect is particularly pronounced under short-prompt conditions, where weight loading is the key bottleneck, \ie under LAMBADA and WinoGrande datasets.

\noindent\textbf{+ Synergistic Granular Pipeline.}
Introducing the synergistic granular pipeline further reduces the idle time and improves workload balance of the prefill computation during cold starts.
This reduction is particularly pronounced on longer datasets.
Specifically, for datasets with long prompts, \ie MMLU and HellaSwag, the average TTFT decreases by 12.7\%.
Meanwhile, for datasets with short prompts, the TTFT also decreases by an average of 3.5\%.

\subsection{In-Depth Analyses}\label{sec:analyses}

\noindent
We further conduct in-depth analyses to evaluate the effectiveness of the three key design components of \EdgeFlow.

\begin{figure*}[!t]
    \centering
    \includegraphics[width=\textwidth]{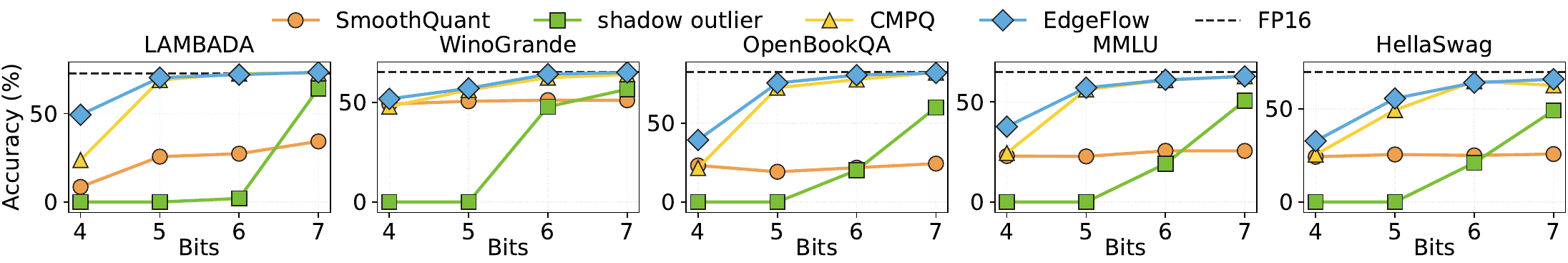}
    \vspace{-7mm}
    \caption{Accuracy of quantization schemes across precisions on Llama3 8B. The horizontal dotted line represents the FP16 accuracy.}
    \label{fig:eval1-quant}
    \vspace{-0.1in}
\end{figure*}
\begin{table}[t]
\centering
\caption{Quantized perplexities on WikiText-2 for Llama3-8B (FP16 perplexity: 14.59). The bold is the best. "SO" means shadow outlier.}
\label{tab:ppl-llama3-wikitext}
\footnotesize
\resizebox{0.48\textwidth}{!}{
\begin{tabular}{c|cccc}
    \toprule
    \textbf{Llama3-8B} & \textbf{SmoothQuant} & \textbf{SO} & \textbf{CMPQ} & \textbf{\EdgeFlow} \\
    \midrule
        4bits & 410.67 & 10k+ & 103.52 & \textbf{30.96} \\
        5bits & 99.68 & 10k+ & 18.22 & \textbf{17.27} \\
        6bits & 97.31 & 814.51 & 15.83 & \textbf{15.63} \\
        7bits & 75.00 & 21.62 & 15.11 & \textbf{15.09} \\
    \bottomrule
\end{tabular}}
\vspace{-0.2in}
\end{table}

\subsubsection{Analyses of NPU-Aware Adaptive Quantization}\label{sec:eval1}
\noindent
We evaluate the accuracy and perplexity of the NPU-aware adaptive quantization against three state-of-the-art quantization methods, \ie SmoothQuant~\cite{icml23smoothquant}, llm.npu's shadow outlier~\cite{asplos25mllm}, and CMPQ~\cite{arxiv25cmpq}.
We adapt SmoothQuant and CMPQ to comply with NPU constraints.
For SmoothQuant, we apply the NPU smoothing technique to satisfy the NPU's per-tensor output quantization constraint.
For CMPQ, we modify its precision allocation strategy from input-channel-wise to output-channel-wise, and replace its allocation metric with our relative error accordingly.

Tables~\ref{tab:ppl-llama3-wikitext} report perplexities of quantized models in the Wikitext-2~\cite{iclr17wikitext2} dataset.
\EdgeFlow steadily yields lower perplexities than all baselines in various precisions.
Figure~\ref{fig:eval1-quant} shows the accuracy in different datasets and integer precisions for Llama3 8B.
As the results across different models are similar, the remaining results are provided in Appendix~\ref{appendix:appendix_quant_eval}.

\EdgeFlow consistently outperforms previous methods across all settings.
Specifically, compared with SmoothQuant, shadow outlier, and CMPQ, respectively, \EdgeFlow achieves average accuracy improvements of 16.55\%–38.23\%, 13.72\%–63.13\%, and 0.67\%–13.55\% across 4 to 7 bits.
Notably, at lower precisions, \ie 4 and 5 bits, SmoothQuant and shadow outlier exhibit significant accuracy degradation.
This indicates that treating all weights equally fails to capture the importance of different channels.
Moreover, \EdgeFlow exceeds CMPQ by 13.55\%, 2.47\%, 0.67\%, and 0.90\% at 4 to 7 bits, showing the effectiveness of our precision assignment strategy.

\begin{figure}[!t]
    \centering
    \begin{tikzpicture}
        \node[anchor=south west,inner sep=0] (img)
            {\includegraphics[width=1\columnwidth]{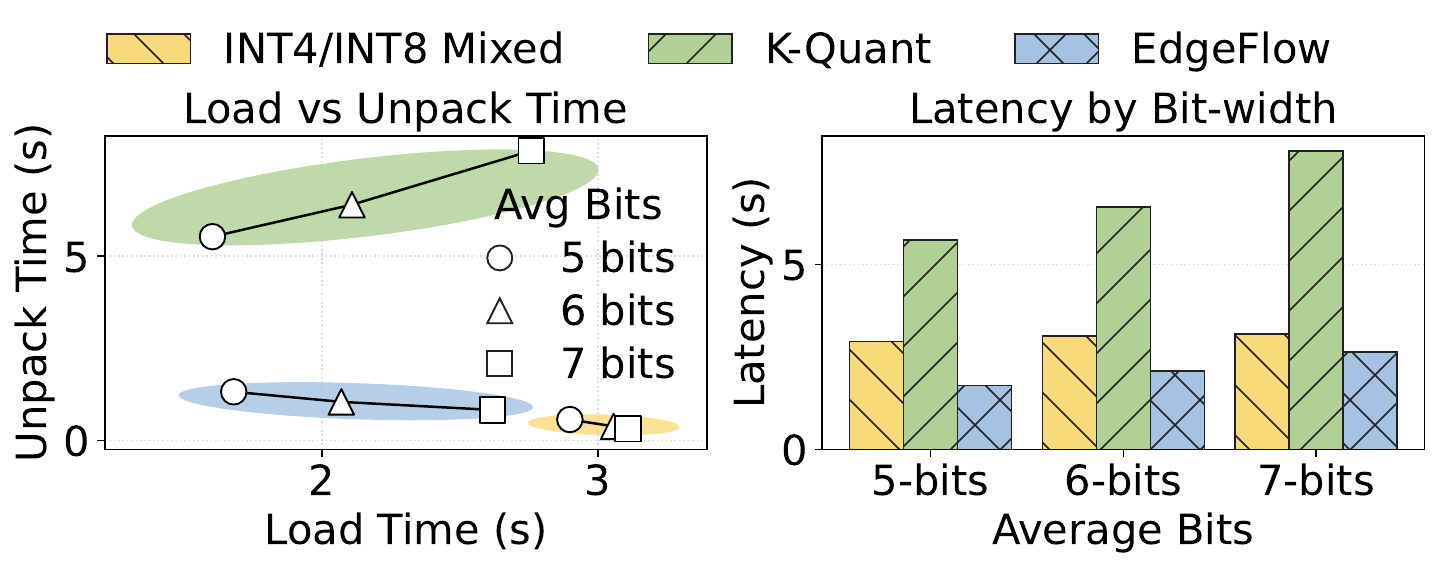}};
        \begin{scope}[overlay]
        \node[anchor=north] at
            ($(img.south west)!0.27!(img.south east)$)
            {\small{(a) Loading and unpacking times.}};
        \phantomsubcaption
        \label{fig:eval2-unpack:a}
        \node[anchor=north] at
            ($(img.south west)!0.78!(img.south east)$)
            {\small{(b) The end-to-end latency.}};
        \phantomsubcaption
        \label{fig:eval2-unpack:b}
        \end{scope}
        \path (img.south west) ++(0,-2mm);\textbf{}
    \end{tikzpicture}
    \caption{
        Performance comparison of different storage formats.
    }
    \label{fig:eval2-unpack}
    \vspace{-0.15in}
\end{figure}
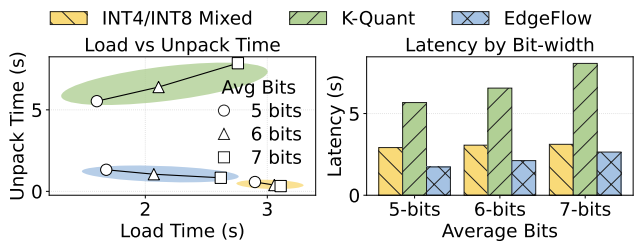

\subsubsection{Analyses of SIMD-Friendly Packing Format}\label{sec:eval2}
\noindent
We evaluate the performance of \EdgeFlow's SIMD-friendly weight packing format against an INT4/INT8 mixed storage format and the K-Quant format.
We use Llama3-8B and quantize it into varying bit-widths, \ie 5, 6, and 7 bits.

Figure~\ref{fig:eval2-unpack:a} shows the unpacking and weight loading times of the three approaches.
Compared with the INT4/INT8 mixed format, \EdgeFlow speeds up model loading by $1.42\times$ since it eliminates read amplification incurred by padding weights.
The additional overhead on unpacking is less than 0.65s due to our efficient SIMD-based unpacking algorithm.
Compared with the K-Quant format, \EdgeFlow has identical load times since both approaches store weights in a compact format.
However, the unpacking time is reduced by $6.17\times$ thanks to the SIMD-based unpacking.

Figure~\ref{fig:eval2-unpack:b} presents the end-to-end latency under the same configuration.
\EdgeFlow consistently outperforms baselines across all bit-widths, achieving an average $3.05\times$ and $1.44\times$ latency reduction compared with K-Quant and INT4/INT8 mixed formats, respectively.
Both results indicate that our SIMD-friendly storage format achieves a better trade-off between read amplification and unpacking efficiency.

\subsubsection{Analyses of Synergistic Granular Pipeline}\label{sec:eval3}
\noindent
We analyze the pipeline of \EdgeFlow to show the effectiveness of the fine-grained operator placement, the position-guided priority, and the task stealing scheme.
We use a sequence length of 512 for Llama3 8B and Phi3 3.8B.
The baseline is the static and coarse-grained pipeline of llm.npu.
\textsf{+~Place} adopts the fine-grained operator placement scheme without dynamic scheduling.
\textsf{+~Priority} incorporates the position-guided priority and \textsf{+~Steal} enables CPU task stealing.

\begin{figure}[!t]
    \centering
    \begin{tikzpicture}
        \node[anchor=south west,inner sep=0] (img)
            {\hspace{-3mm}\includegraphics[width=1\columnwidth]{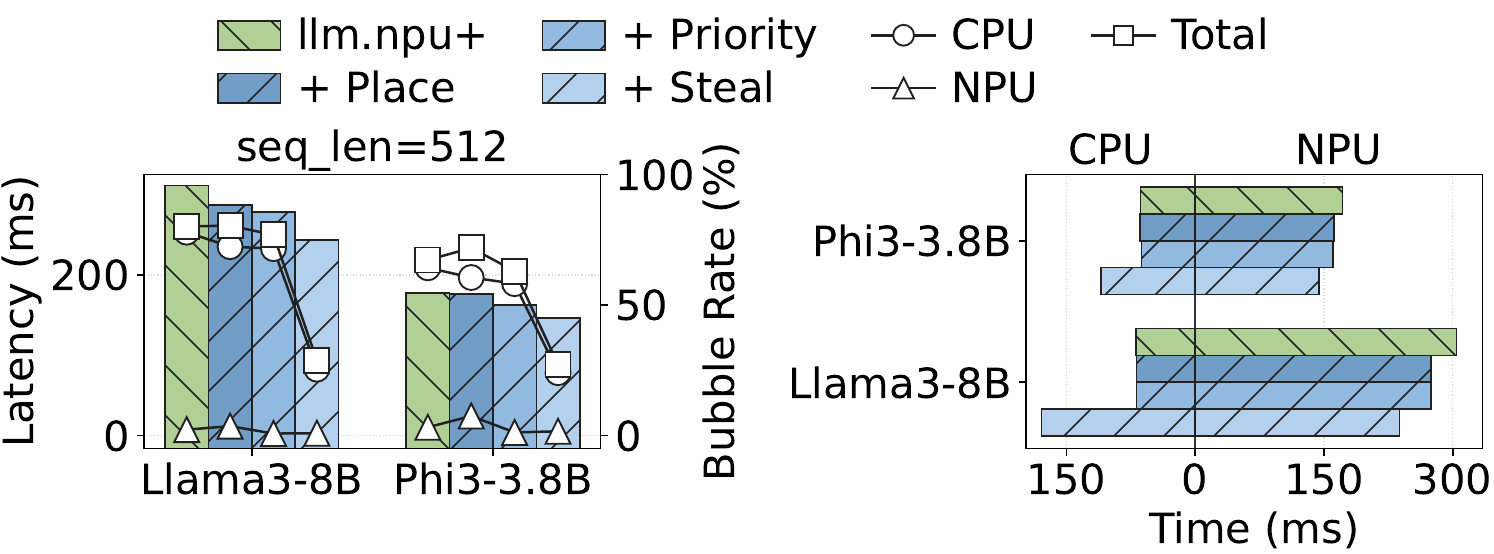}};
        \begin{scope}[overlay]
        \node[anchor=north, text width=0.55\columnwidth, align=center] at
            ($(img.south west)!0.23!(img.south east)$)
            {\small{(a) Single-layer execution times and bubble rates.}};
        \phantomsubcaption
        \label{fig:eval3-schedule:a}
        \node[anchor=north, text width=0.45\columnwidth, align=center] at
            ($(img.south west)!0.75!(img.south east)$)
            {\small{(b) Single-layer execution times of the NPU and CPU.}};
        \phantomsubcaption
        \label{fig:eval3-schedule:b}
        \end{scope}
        \path (img.south west) ++(0,-6mm);
    \end{tikzpicture}
    \caption{
        Analyses of the synergistic granular pipeline.
    }
    \label{fig:eval3-schedule}
    \vspace{-0.15in}
\end{figure}
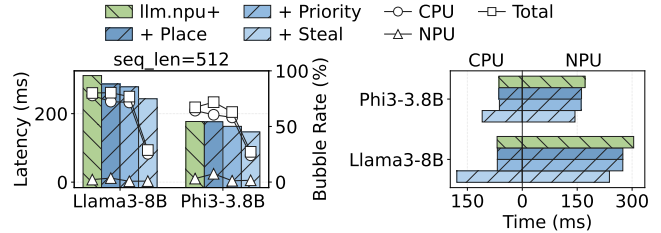

Figure~\ref{fig:eval3-schedule:a} shows the single-layer execution latency and the bubble rates of the CPU and NPU.
Figure~\ref{fig:eval3-schedule:b} presents the execution times of CPU and NPU.
Compared with llm.npu, \textsf{+ Place} decreases total latency by 4.2\%, and NPU execution time is reduced by 7.7\% with nearly unchanged CPU time.
This improvement mainly arises from offloading NPU-inefficient operators to the CPU, which distributes the workload more appropriately.
However, \textsf{+ Place} employs a static CPU–NPU scheduling, which almost doubles the NPU bubble rate and leads to significant idle periods.

\textsf{+ Priority} further reduces latency by 5.2\% and lowers the NPU bubble rate by 82.2\%.
This is achieved by ordering operator executions according to their positions in the prompt, allowing downstream NPU tasks to be unlocked sooner.
Enabling task stealing, \ie \textsf{+ Steal}, further decreases execution time by 11.3\% and reduces the CPU bubble rate by 61.6\%.
The CPU and NPU execution time also becomes more balanced. 
This gain results from better CPU utilization through opportunistic execution of NPU-assigned tasks.
In particular, the NPU bubble rate stays roughly unchanged.
This is attributed to the task-stealing threshold, which avoids task-stealing becoming so aggressive that it could make the NPU idle.

\subsection{Deployment Efficiency}\label{sec:overhead}

\noindent
We finally compare \EdgeFlow with llm.npu in terms of decoding efficiency and resource consumption.

\subsubsection{Decoding Efficiency}\label{eval4-decoding}
\begin{figure}[!t]
    \vspace{1mm}
    \centering
    \includegraphics[width=1\columnwidth]{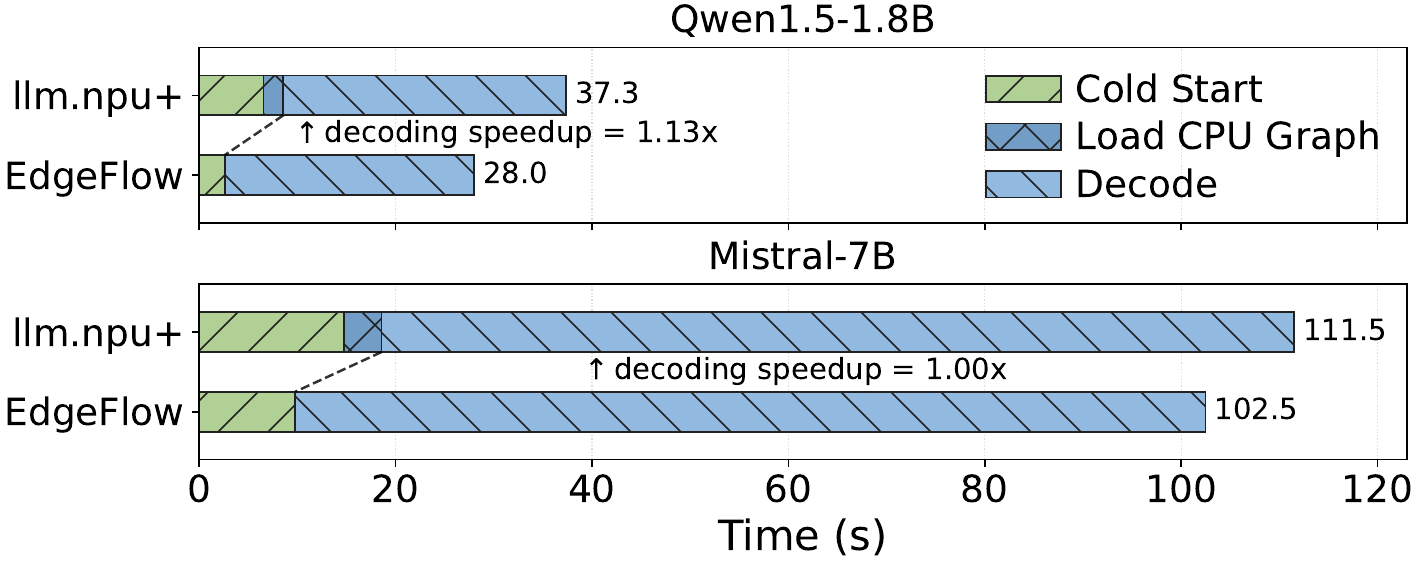}
    \vspace{-8mm}
    \caption{Breakdown of end-to-end completion latency.}
    \label{fig:eval4-decoding}
    \vspace{-0.2in}
\end{figure}

\noindent
We break down the end-to-end completion process of Mistral 7B and Qwen1.5 1.8B (512 input tokens, 512 generated tokens) on \EdgeFlow and llm.npu, as shown in Figure~\ref{fig:eval4-decoding}.
Since llm.npu performs decoding on the CPU, we adopt the same setup for \EdgeFlow to ensure a fair comparison.
llm.npu adds a 2.9s overhead between cold start and decoding when switching to a CPU decoding graph, while \EdgeFlow avoids this transition and enters decoding seamlessly.
Isolating the impact on decoding itself, \EdgeFlow achieves a speedup of 1.13$\times$ on Qwen1.5 1.8B and 1.00$\times$ on Mistral 7B compared with llm.npu.
This shows that despite introducing additional processing during cold start, \EdgeFlow does not degrade decoding efficiency.
Instead, it preserves comparable decoding performance while reducing overall latency through better cold start design.

\subsubsection{Resource Consumption}\label{eval4-resource}
\begin{figure}[!t]
    \centering
    \begin{tikzpicture}
        \node[anchor=south west,inner sep=0] (img)
            {\hspace{-3mm}\includegraphics[width=0.95\columnwidth]{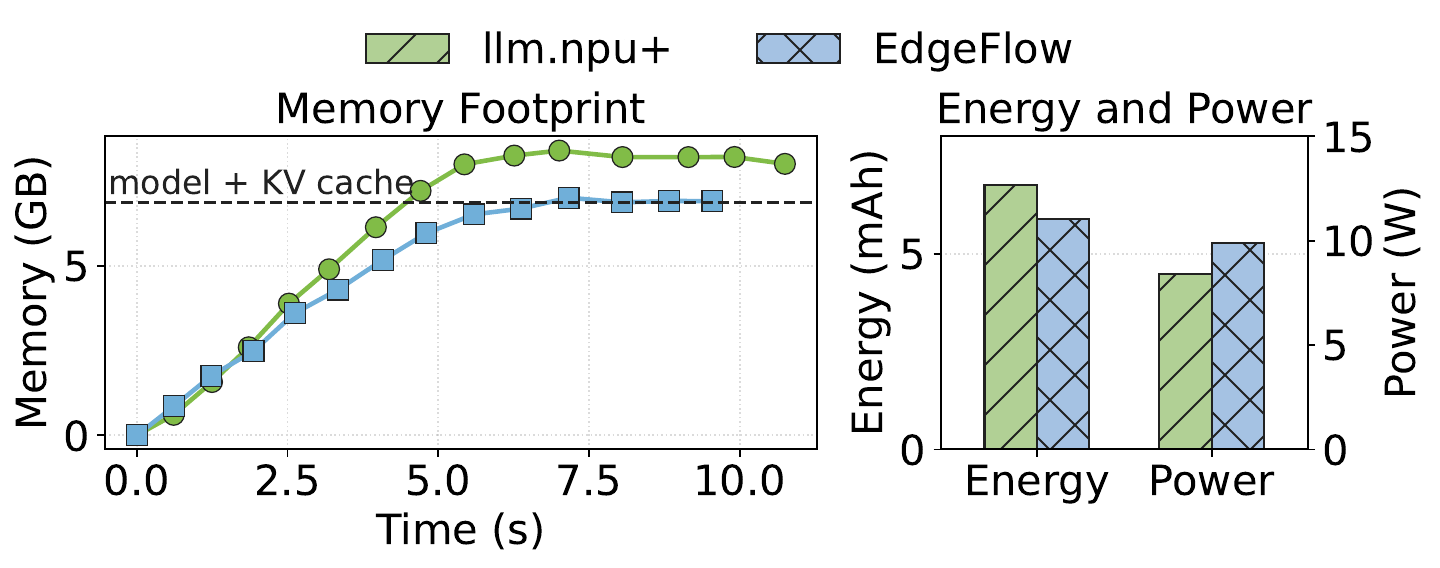}};
        \begin{scope}[overlay]
        \node[anchor=north, text width=0.53\columnwidth, align=center] at
            ($(img.south west)!0.27!(img.south east)$)
            {\small{(a) Memory footprint.}};
        \phantomsubcaption
        \label{fig:eval4-resource:a}
        \node[anchor=north, text width=0.48\columnwidth, align=center] at
            ($(img.south west)!0.77!(img.south east)$)
            {\small{(b) Energy and power.}};
        \phantomsubcaption
        \label{fig:eval4-resource:b}
        \end{scope}
        \path (img.south west) ++(0,-2mm);
    \end{tikzpicture}
    \caption{
        Comparison of resource consumption during the cold start phase of  Mistral 7B with 512 tokens.
    }
    \label{fig:eval3-resource}
    \vspace{-0.20in}
\end{figure}
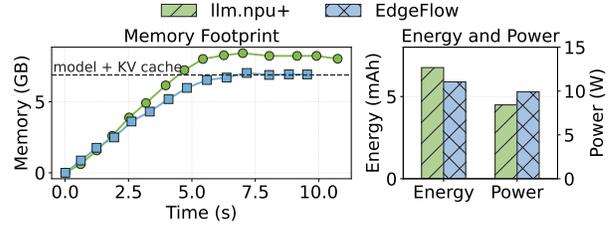

Figure~\ref{fig:eval4-resource:a} shows the memory footprint of \EdgeFlow and llm.npu during cold start.
Ideally, the memory footprint of both approaches should be the summation of model size under INT8 quantization and the KV cache size ($\approx 6.9$ GB in total) since both approaches need to dequantize model weights to INT8 before computation.
Initially, the memory usage of both approaches increases gradually since weights are loaded in a layer-by-layer manner.
When it comes to the stable stage, llm.npu has higher memory consumption ($\approx$ 8 GB) than the ideal case since it maintains separate computation graphs for CPU and NPU.
The footprint of \EdgeFlow is close to ideal since the weights in the computation graph are shared among CPU and NPU.

Figure~\ref{fig:eval4-resource:b} reports the power and energy consumption during the same cold start process.
\EdgeFlow exhibits a 17.6\% higher average power than llm.npu.
This is because \EdgeFlow offloads more lightweight operators to the CPU and opportunistically steals operators from the NPU.
Since CPU execution generally incurs higher power than NPU, this leads to increased average power.
Nevertheless, \EdgeFlow achieves lower total energy consumption, \ie reducing by 12.9\%.
This is due to reduced redundant data loading from flash storage and a more efficient execution pipeline, resulting in a shorter cold start time.

\section{Related Work}\label{sec:related_work}

\noindent
\textbf{LLM Inference on Mobile Devices.}
Existing works on mobile LLM inference can be categorized into computation and storage optimizations.
Computation-oriented approaches improve the execution efficiency of the prefill computation.
llm.npu~\cite{asplos25mllm} executes attention on the CPU while running the remaining parts on the NPU.
HeteroInfer~\cite{sosp25heteroinfer} places dense kernels on the NPU and offloads remaining operations to the GPU based on NPU sensitivity profiling.
Moreover, ShadowAttn~\cite{arxiv25shadowattn} approximates important tokens in low-precision on the NPU and performs high-precision sparse attention on the CPU.
Storage-oriented approaches offload weights and activations on the flash to execute larger models and reduce I/O overhead during offloading.
LLM in a Flash~\cite{acl24llmflash} bundles co-activated FFN neurons to avoid fragmented flash reads.
ELMS~\cite{mobicom25elms} permutes weights by importance so that latency-critical prefixes can be loaded consecutively.
Finally, PowerInfer-2~\cite{arxiv24powerinfer2} keeps hot FFN neurons resident in memory and loads cold neurons on demand.

Compared with existing approaches, \EdgeFlow is the first to optimize cold inferences for mobile LLMs.
Moreover, techniques proposed in \EdgeFlow are also complementary to existing approaches.
Specifically, the NPU-aware adaptive quantization algorithm and the SIMD-friendly packing format can be adopted together with existing storage-oriented works to further reduce I/O overhead during model offloading.
The synergistic granular pipeline further improves the compute efficiency of existing compute-oriented approaches with its dynamic and fine-grained design.

\noindent\textbf{Mixed-Precision LLM Quantization.}
These algorithms leverage the varying importance of model weights by assigning different bit-widths to achieve a better accuracy and efficiency trade-off.
LLM-MQ~\cite{nips23llm-mq} uses first-order loss sensitivity to guide tensor-level precision assignment, while keeping sparse outlier weights in FP16.
SliM-LLM \cite{icml25slim} allocates bits based on reconstruction error, and performs group-wise calibration of quantization parameters.
ResQ~\cite{icml25resq} uses principal component analysis to identify high-variance subspaces and performs rotation to mitigate outliers.
CMPQ~\cite{arxiv25cmpq} profiles per-channel input magnitudes to estimate weight importance, and assigns bit-widths heuristically with non-uniform quantization.
Compared with these approaches, \EdgeFlow accounts for NPU-specific quantization constraints and delivers superior model accuracy under these constraints.

\noindent\textbf{Cold Start for Model Serving Systems.}
Existing cold-start mitigation techniques can be classified into cloud-based and mobile-based techniques.
Cold-start mitigation techniques on the cloud are designed for GPU-based inference systems.
Specifically, ServerlessLLM~\cite{osdi24serverlessllm} leverages a multi-tier checkpoint caching and a GPU-aware formatting to reduce the provisioning latency of an inference instance.  
Medusa~\cite{asplos25medusa} and PhoenixOS~\cite{sosp25pos} materialize and load GPU execution states to enable faster initialization.  
ParaServe~\cite{arxiv25paraserve} and BLITZSCALE~\cite{osdi25blitz} aggregate bandwidth across GPU servers to parallelize model loading.
Compared with these approaches, \EdgeFlow tackles cold starts for mobile LLM inference, which cannot adopt these GPU-based approaches.

For mobile devices, NNV12~\cite{mlsys23nnv12} reduces cold-start latency by selecting startup-friendly kernels, which is complementary to \EdgeFlow.
There are also OS-level optimizations to reduce cold start latency for general applications, which can be broadly categorized into data prefetching~\cite{fast11fast,atc21asap,mobicom26appflow}, memory paging and swapping management~\cite{hpca25ariadne,fast25archer}, and faster memory reclamation~\cite{atc20acclaim,atc25prm}.
These works are orthogonal to \EdgeFlow since we focus on reducing the pure cold-start latency with algorithm and system co-design.
\vspace{-0.15in}
\section{Conclusion}
\noindent
In this paper, we identify that the inefficient flash bandwidth utilization significantly deteriorates the cold start latency of mobile LLM inferences.
We design \EdgeFlow and show that the cold start latency can be mitigated by assigning precisions to weights adaptively according to their importance.
\EdgeFlow achieves this idea with three techniques, \ie an NPU-aware adaptive quantization, an SIMD-friendly packing format, and a synergistic granular pipeline. 
Our evaluation results show that \EdgeFlow significantly reduces cold-start latency up to $4.07\times$ compared with state-of-the-art baselines while maintaining comparable model accuracy.

\bibliographystyle{ACM-Reference-Format}
\bibliography{main}

\clearpage
\appendix

\section{Relative Error Derivation}
\label{appendix:relative_error_derivation}

\noindent
We derive the relationship between the cosine distance and the expected squared quantization error for weight vectors. Let $W \in \mathbb{R}^n$ denote a weight vector and $W'$ its quantized-and-dequantized version. We define the quantization error vector as a random variable.
\[
E := W - W', \quad E = (E_1, \dots, E_n),
\]
where each $E_i$ represents the quantization error of a component of $W$.

The cosine similarity between $W$ and $W'$ is
\[
\cos \theta = \frac{W \cdot W'}{\|W\| \, \|W'\|}.
\]

We analyze this by substituting $W' = W - E$:
\begin{enumerate}
    \item First, we analyze the numerator: $$W \cdot W' = W \cdot (W - E) = W \cdot W - W \cdot E = \|W\|^2 - W \cdot E$$
    \item Second, we analyze the norm of $W'$ in the denominator:$$\|W'\|^2 = (W - E) \cdot (W - E) = \|W\|^2 - 2W \cdot E + \|E\|^2$$
\end{enumerate}

To simplify, we adopt two standard assumptions in quantization analysis:

\begin{enumerate}
    \item \textbf{Small error energy}: $\|E\|^2 \ll \|W\|^2$.
    \item \textbf{Uncorrelated error}: $E$ is approximately uncorrelated with $W$, i.e., $\mathbb{E}[W \cdot E] \approx 0$.
\end{enumerate}

Under these assumptions, the denominator can be expanded using a first-order Taylor approximation:
\begin{align*}
\|W'\| 
&= \sqrt{\|W\|^2 - 2 W \cdot E + \|E\|^2} \\[2mm]
&\approx \sqrt{\|W\|^2 + \|E\|^2} \quad (\text{since } W \cdot E \approx 0) \\[1mm]
&= \|W\| \sqrt{1 + \frac{\|E\|^2}{\|W\|^2}} \\[1mm]
&\approx \|W\| \left( 1 + \frac{\|E\|^2}{2 \|W\|^2} \right) \quad (\text{for small } \|E\|^2).
\end{align*}

Substituting back, the cosine similarity becomes
\[
\cos \theta \approx \frac{\|W\|^2}{\|W\|^2 \left(1 + \frac{\|E\|^2}{2 \|W\|^2}\right)}
= \frac{1}{1 + \frac{\|E\|^2}{2 \|W\|^2}}
\approx 1 - \frac{\|E\|^2}{2 \|W\|^2}.
\]

Since $E_i$ are modeled as i.i.d.\ random variables with zero mean and finite variance, we can connect the squared L2 norm to the expected squared error:
\[
\mathbb{E}[\|E\|^2] = \mathbb{E}\Big[\sum_{i=1}^n E_i^2\Big] = \sum_{i=1}^n \mathbb{E}[E_i^2] = n \cdot \mathbb{E}[E^2].
\]
Similarly, the L2 norm of the weights can be written in terms of the mean squared value:
\[
\|W\|^2 = \sum_{i=1}^n W_i^2 = n \cdot \mathbb{E}[W^2],
\]
where $\mathbb{E}[W^2]$ denotes the mean square of the weight components.

Finally, substituting these into the cosine distance approximation, we obtain
\[
1 - \cos \theta \approx \frac{\mathbb{E}[\|E\|^2]}{2 \, \|W\|^2} \approx \frac{n \cdot \mathbb{E}[E^2]}{2 \, (n \cdot \mathbb{E}[W^2])} = \frac{\mathbb{E}[E^2]}{2 \, \mathbb{E}[W^2]}.
\]
This justifies the use of the expected squared error as a tractable approximation for the cosine distance between a weight vector and its quantized version.

\begin{figure}[t]
    \centering
    \includegraphics[width=0.85\columnwidth]{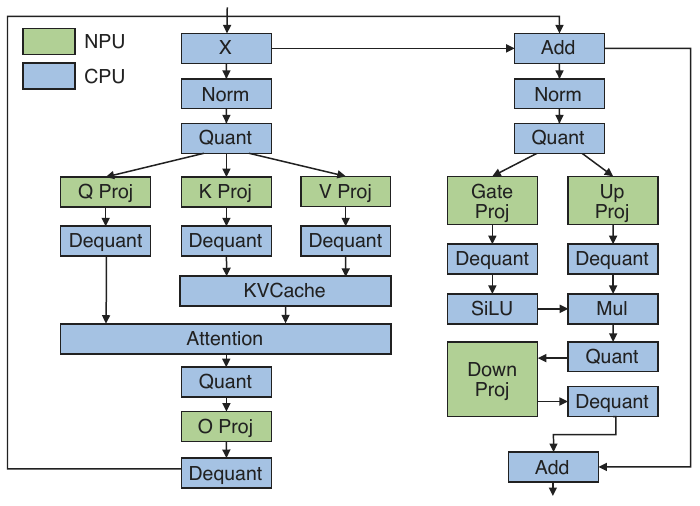}
    \caption{Transformer layer operator placement: CPU vs. NPU.}
    \label{fig:transformer_layer}
\end{figure}

\begin{figure*}[t]
    \centering
    \includegraphics[width=\textwidth]{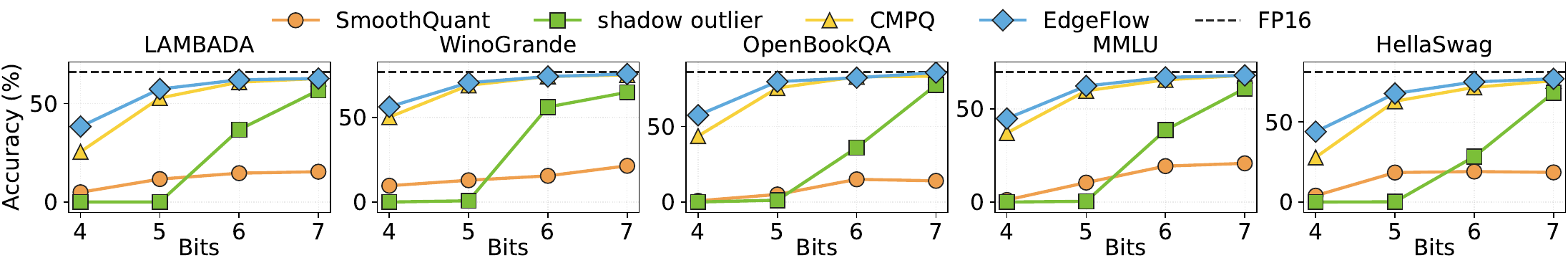}
    \caption{Accuracy of quantization schemes across precisions on Phi3 3.8B. The horizontal dotted line represents the FP16 accuracy.}
    \label{fig:eval-phi3-quant}
\end{figure*}

\section{Operator-Level Placement}
\label{appendix:operator_level_placement}
We provide a detailed operator-level execution flow of a transformer layer under \EdgeFlow, including device placement, precision transitions, and data dependencies.

\EdgeFlow assigns all INT8 matrix multiplications to the NPU, including the Q/K/V/O projections in attention and the Gate/Up/Down projections in feedforward networks (FFNs).
This design is motivated by the architecture of the Hexagon NPU, which is equipped with dedicated HMX units optimized for high-throughput matrix operations.
In particular, INT8 matrix multiplications achieve significantly higher throughput compared to FP16 on the NPU.

In contrast, the remaining operators are predominantly element-wise or low arithmetic-intensity operations, such as normalization, activation functions, and residual additions.
Offloading these operators to the NPU would not yield meaningful speedup due to limited hardware specialization and potential overheads. Therefore, they are executed on the CPU in FP16 precision, which provides better efficiency for such workloads.

As illustrated in Figure~\ref{fig:transformer_layer}, each block represents an individual operator that is scheduled as a task and inserted into either the CPU or NPU task queue based on its placement.
The annotated edges form a topological backbone that captures the data dependencies among operators within a transformer layer.
During execution, operators are dispatched in a dependency-aware manner.
While respecting the topological order, \EdgeFlow further optimizes execution using position-guided priority scheduling and task stealing, enabling dynamic load balancing across devices.

\begin{table}[t]
\centering
\caption{Quantized perplexities on WikiText-2 for Phi3-3.8B (FP16 perplexity: 10.01). The bold is the best. "SO" means shadow outlier.}
\label{tab:ppl-phi3-wikitext}
\footnotesize
\resizebox{0.48\textwidth}{!}{
\begin{tabular}{c|cccc}
    \toprule
    \textbf{Phi3-3.8B} & \textbf{SmoothQuant} & \textbf{SO} & \textbf{CMPQ} & \textbf{\EdgeFlow} \\
    \midrule
        4bits & 338.62 & 10k+ & 35.87 & \textbf{25.88} \\
        5bits & 93.34 & 10k+ & 15.54 & \textbf{14.24} \\
        6bits & 77.49 & 42.17 & 11.90 & \textbf{11.65} \\
        7bits & 77.27 & 13.82 & 10.98 & \textbf{10.93} \\
    \bottomrule
\end{tabular}}
\end{table}

\section{Adaptive Quantization Evaluation}
\label{appendix:appendix_quant_eval}
We further evaluate adaptive quantization on Phi3 3.8B, with detailed results reported in this section.
Similar to Llama3 8B, Table~\ref{tab:ppl-phi3-wikitext} presents the perplexities on the Wikitext-2~\cite{iclr17wikitext2} dataset, where \EdgeFlow consistently achieves lower perplexities than all baselines across different precisions.
Figure~\ref{fig:eval-phi3-quant} illustrates the accuracy under various integer precisions on multiple datasets.

Overall, \EdgeFlow maintains clear advantages over prior methods in all evaluated settings.
Compared with SmoothQuant, shadow outlier, and CMPQ, \EdgeFlow improves accuracy by 44.07\%–55.87\%, 8.15\%–67.06\%, and 0.82\%–11.29\%, respectively, across 4 to 7 bits.
Consistent with observations on Llama3, SmoothQuant and shadow outlier suffer from notable accuracy degradation at lower precisions (e.g., 4–5 bits), suggesting that uniform treatment of weights is insufficient to capture channel-wise importance.

In contrast, \EdgeFlow demonstrates stable performance across all bit-widths, outperforming CMPQ by 11.29\%, 3.54\%, 1.05\%, and 0.82\% at 4 to 7 bits, respectively.
These results further validate the effectiveness of our adaptive precision assignment strategy and demonstrate that the benefits of \EdgeFlow generalize well across different model architectures.

\end{document}